\documentclass[11pt]{article}
\usepackage{openwork}
\usepackage{graphicx}
\usepackage{subcaption}
\usepackage{xcolor}
\usepackage{float}
\usepackage{placeins}
\usepackage{textgreek}
\usepackage{booktabs}
\usepackage{multirow}
\usepackage{graphicx}

\title{PDO Drug Screening Data for Evaluating and Training AI Drug Response Prediction Models}
\author[1,$*$]{Yitan Zhu}
\author[1]{Kaitlyn Marlor}
\author[1]{Vincent Lavelle}
\author[2]{Thomas S Brettin}
\author[3]{Matteo Sacco}
\author[3]{Costanza Siniscalchi}
\author[4]{Henry W Raeder}
\author[5]{Michael Ryan Weil}
\author[2, 6]{Rick L Stevens}
\affil[1]{Division of Data Science and Learning, Argonne National Laboratory, Lemont, IL 60439, USA}
\affil[2]{Computing, Environment and Life Sciences, Argonne National Laboratory, Lemont, IL 60439, USA}
\affil[3]{Department of Medicine, Section of Hematology/Oncology, University of Chicago, Chicago, IL, USA}
\affil[4]{Department of Human Genetics, University of Chicago, Chicago, IL 60637, USA}
\affil[5]{Cancer Research Technology Program, Cancer Data Science Initiatives, Frederick National Laboratory for Cancer Research, Frederick, MD 21701, USA}
\affil[6]{Department of Computer Science, The University of Chicago, Chicago, IL 60637, USA}
\affil[$*$]{Correspondence: yitan.zhu@anl.gov}

\date{}

\makeatletter
\renewcommand{\@author}{%
  \AB@authlist\par\vspace{\affilsep}%
  \begingroup
  \raggedright
  \AB@affillist\par
  \endgroup
}
\makeatother

\begin{document}

\maketitle

\vspace{1.5em}

\begin{abstract}
Anti-cancer drug response prediction (DRP) is critically important for precision oncology and therapeutic development. Currently, most DRP models are developed using drug screening data from immortalized cancer cell lines, which can differ substantially from patient tumors at the genomic and molecular levels, even though patient tumor response is the ultimate prediction target. To help bridge this gap and bring DRP models closer to their intended clinical application, we conducted four drug screening studies using patient-derived organoids (PDOs), which more faithfully recapitulate patient tumors than immortalized cell lines. These studies generated data for improving and evaluating DRP models, including multi-omics profiles of PDO lines, drug feature data, and treatment response measurements. Using these data, we retrained and evaluated multiple publicly available DRP models under three model-building scenarios: training models using cell line data only, training models using combined cell line and PDO data, and pretraining models on cell line data followed by fine-tuning on PDO data. We then evaluated and compared model prediction performance on PDOs across these analysis scenarios. In addition, we also assessed the consistency of drug screening viability data generated by different laboratories using different assay technologies and/or experimental protocols.

\end{abstract}

\vspace{1.5em}

\section {1 Introduction}

Cancer remains a major global health challenge despite continued advances in diagnosis and treatment. Precision oncology seeks to improve therapeutic outcomes by using patient-specific molecular characteristics to guide treatment selection, and the increasing availability of genomic and other molecular profiling data has created new opportunities for computational approaches to support this goal \cite{Partin2023}. In particular, artificial intelligence (AI) and machine learning methods are increasingly being developed for drug response prediction (DRP), in which molecular features of a cancer case and characteristics of a therapeutic compound are used to predict treatment sensitivity or resistance. A wide range of approaches, including graph neural networks \cite{Nguyen2021}, transformers \cite{jiang2022deeptta}, and gradient-boosted decision trees \cite{lgbm}, have been developed for this task. However, the utility of these models depends fundamentally on the experimental data available for their development and evaluation.

Most existing DRP models have been developed using large-scale pharmacogenomic datasets generated using immortalized cancer cell lines (CCLs), including resources such as the Cancer Cell Line Encyclopedia (CCLE) \cite{Barretina2012} and Genomics of Drug Sensitivity in Cancer (GDSC) \cite{Yang2013}. CCLs are relatively inexpensive to culture, experimentally reproducible, and capable of indefinite proliferation, making it possible to generate drug response measurements across large numbers of cancer models and compounds. These properties have made CCL datasets an essential resource for the development of computational DRP methods. However, prolonged in vitro culture and growth in two-dimensional monolayers can alter molecular characteristics \cite{BenDavid2018} and fail to reproduce important features of patient tumors, including cellular heterogeneity, three-dimensional tissue architecture, and aspects of the tumor microenvironment. This creates an important gap between the experimental systems that provide most of the data used to train DRP models and the patient tumors to which such models are ultimately intended to be applied. Patient-derived organoids (PDOs) offer an experimental system that can help bridge the gap between conventional cancer cell lines and patient tumors. PDOs are three-dimensional cultures established from patient tumor tissues and can retain genomic, molecular, and phenotypic characteristics of the original tumors more faithfully than immortalized cell lines. These properties make PDOs a promising preclinical model for evaluating therapeutic response and for developing computational models that more closely reflect the intended application of precision oncology. However, PDO pharmacogenomic data remain limited compared with the large-scale drug screening data available for CCLs, posing a significant barrier to the development and evaluation of DRP models for PDOs.

In this work, we address this need by generating new PDO drug screening data specifically designed to support the development and evaluation of AI-based DRP models. We conducted four drug screening studies spanning colorectal cancer, head and neck cancer, and sarcoma PDOs, producing matched drug response measurements and molecular profiles across multiple experimental platforms and assay protocols. Three of these studies were designed with overlapping PDOs and compounds, enabling direct comparison of drug response measurements generated independently using different screening technologies and experimental procedures. Together, these studies provide new data for investigating both the reproducibility of PDO drug screening measurements and the ability of computational models to predict drug response in experimental systems that more closely represent patient tumors. 

We further use these newly generated datasets to investigate how PDO data can support the development and evaluation of DRP models. Six publicly available DRP models representing diverse machine learning and deep learning architectures were evaluated under three learning strategies. The first strategy trains models exclusively on large-scale cancer cell line (CCL) data and directly applies them to predict drug responses in PDOs. The second strategy trains models on a combination of CCL data and a subset of PDO data, with predictive performance evaluated on the remaining PDO data. The third strategy pretrains models on CCL data and subsequently fine-tunes them using a subset of PDO data, with performance evaluated on the remaining PDO data. Model performance was assessed using two cross-validation schemes: random partitioning, in which data are randomly split into training, validation, and test sets, and cancer-blind partitioning, in which cancer cases are mutually exclusive across these sets. The cancer-blind scheme is designed to assess model generalizability to previously unseen cancer cases. By systematically evaluating multiple model architectures and learning strategies across independently generated PDO datasets, we characterize the extent to which knowledge learned from large-scale CCL datasets transfers to PDOs and whether limited PDO data can improve predictive performance in the target domain. Together, these analyses demonstrate the utility of the newly generated datasets for developing and evaluating DRP models for PDO drug response prediction.

\section{2  Materials and Methods}

\subsection{2.1 Data Curation}

\subsection{2.1.1 PDO Drug Screening Studies}

To bridge the gap between cell lines and patient tumors and to bring DRP models closer to their ultimate goal of predicting patient tumor response, we conducted four drug screening studies on PDOs of colorectal cancers, head and neck cancers, and sarcoma to collect matched treatment response and omics data with multiple drug screening technologies. Three of the drug screening studies were conducted for the Innovative Methodologies and New Data for Predictive Oncology Model Evaluation (IMPROVE) project \cite{Weil2024}, and their datasets are named "IMPROVE dataset 1", "IMPROVE dataset 2", and "IMPROVE dataset 3". The other drug screening study was conducted for our previous Pilot 1 project whose goal was to develop novel effective DRP approaches, and its dataset is named "Pilot 1". See Table~\ref{tab:dataset_summary} for the numbers of PDO/cell lines, compounds, and experiments (pairs of PDO/cell lines and compounds) included in the four datasets.

\begin{table}[ht]
\centering
\caption{Summary of datasets and experimental sources. The final row reports
the total number of unique experiments, drugs, cell lines, and organoids
across all sources.}
\label{tab:dataset_summary}
\resizebox{\textwidth}{!}{%
\begin{tabular}{lrrrr}
\hline
\textbf{Source} &
\textbf{\# Experiments} &
\textbf{\# Drugs} &
\textbf{\# Cell Lines} &
\textbf{\# Organoids} \\
\hline
CCLE                                       & 5,803   & 24  & 470 & 0 \\
CTRPv2                                    & 282,745 & 494 & 803 & 0 \\
GDSCv1                                    & 119,758 & 294 & 619 & 0 \\
GDSCv2                                   & 53,893  & 168 & 539 & 0 \\
gCSI                                     & 3,130   & 16  & 355 & 0 \\
IMPROVE dataset 1                         & 600     & 100 & 2   & 4 \\
IMPROVE dataset 2                         & 280     & 40  & 2   & 5 \\
IMPROVE dataset 3                         & 1,200   & 100 & 2   & 10 \\
Pilot 1                                   & 1,316   & 37  & 0   & 47 \\
\hline
\textbf{Overall} &
\textbf{468,725} &
\textbf{766} &
\textbf{883} &
\textbf{57} \\
\hline
\end{tabular}%
}
\end{table}

\begin{figure}[ht]
    \centering
    \includegraphics[width=\textwidth]{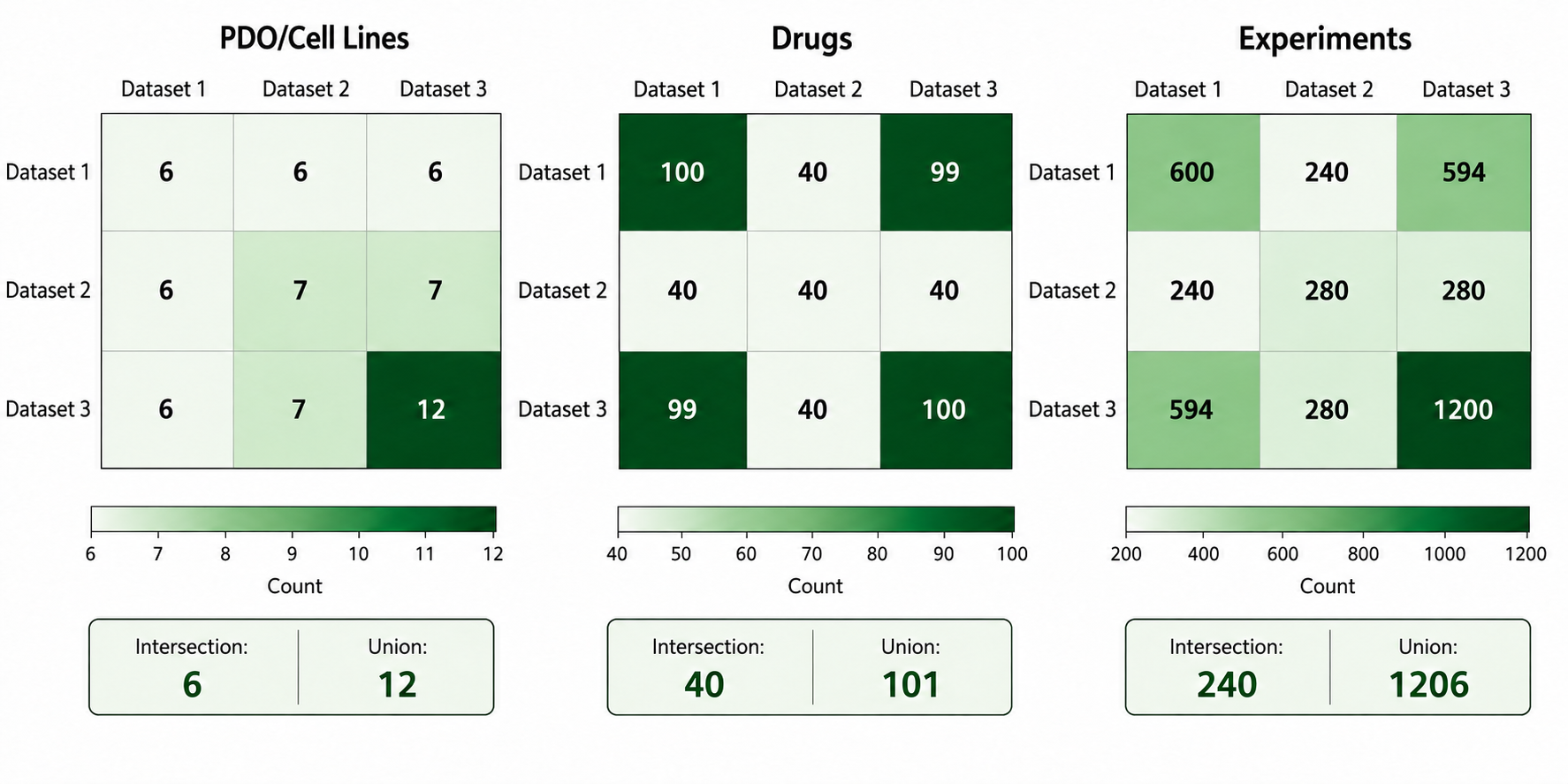}
    \caption{Overlap of samples, experiments, and drugs across the three IMPROVE datasets. 
    Diagonal entries indicate the total number of elements in each set, while 
    off-diagonal entries indicate pairwise intersections. The intersection and 
    union across all three sets are shown below each matrix.}
    \label{fig:improve_overlap}
\end{figure}

IMPROVE drug screening studies 1-3 were designed jointly to share common experiments that enable cross-platform correlation analyses to evaluate the reproducibility of drug screening experiments. IMPROVE datasets 1-3 include 4 colorectal PDO lines, 5 colorectal PDO lines, and 9 colorectal PDO lines plus 1 sarcoma PDO line, respectively. All three IMPROVE drug screening studies also include two common colorectal cancer cell lines for benchmarking purpose. See Figure~\ref{fig:improve_overlap} for the numbers of overlapping PDO/cell lines, drugs, and experiments between studies. The Pilot 1 dataset includes 27 colorectal PDO lines and 20 head and neck cancer PDO lines, which are not included in IMPROVE datasets 1-3. Our drug screening studies focused on colorectal cancer and head and neck cancer, because these two cancer types have lower response prediction accuracies compared with other cancer types based on our previous analysis results. Accumulating more data for these two cancer types helps improve their prediction accuracies. The compounds used in the drug screening studies were selected to cover most of the major mechanisms of actions (MoAs) of current drugs and compounds used in cancer clinical practice and research. The following briefly introduces the experimental protocols that were used to generate the viability measurements in the four screening studies.

\textbf{IMPROVE drug screening study 1} 

Drug responses were evaluated in two colorectal cancer cell lines (HT29 and SW620) and four PDO models (PDM-2, PDM-5, PDM-7, and PDM-9) using a 3D high-content imaging assay. PDO models were developed by the Human Cancer Models Initiative (HCMI) and purchased from the American Type Culture Collection (ATCC). Cell lines were expanded in 2D culture, harvested, mixed with hydrogel, and seeded at 1,250 cells/well into 384-well plates using an automated liquid handler. Organoids were expanded in 3D culture, size-filtered to 20–70 µm, mixed with hydrogel, and similarly seeded into 384-well plates. One day after seeding, compounds were added and cultures were exposed for 5 days. Compounds were evaluated at seven concentrations: 31.6, 3.16, 0.316, 0.0316, 0.00316, 0.000316, and 0.0000316 µM. Staurosporine (10 µM) served as a cytotoxic positive control, while 0.3\% DMSO served as the primary solvent control; 0.3\% DMF or H$_2$O controls were included where appropriate. Conditions were generally tested in triplicate, distributed across replicate plates. After treatment, cultures were fixed with 2\% paraformaldehyde and stained with Rhodamine-Phalloidin and Hoechst to visualize F-actin and nuclei, respectively. 

Z-stack images of each well were acquired. Individual Z-stack planes were first analyzed separately and then the obtained data were integrated across planes to generate numeric features. Key features used in our analyses were organoid count and fraction of living cells. Measurements were normalized per model and plate relative to the appropriate solvent control. Before normalization, outliers were detected from organoid counts per well using the interquartile-range method and removed. Assay quality was evaluated using Z-prime and coefficient of variation (CV) calculated from readouts of control wells, with Z-prime $\geq0.2$ and CV $<20$\% considered acceptable. We used the fraction of living cells to calculate the areas under the dose response curves (AUCs) for the dose range of [100 pM, 100 µM]. AUC values were then normalized by the dose range to obtain normalized AUC values whose typical range was [0, 1].

\textbf{IMPROVE drug screening study 2} 

Drug screening was performed in seven colorectal cancer models, including two cancer cell lines (SW620 and HT-29) and five PDO lines (PDM-2, PDM-4, PDM-5, PDM-7, and PDM-9) developed by the HCMI. Both cell lines and PDO models were purchased from the ATCC. For the cell line screen, cells were embedded in hydrogel, plated, and pre-cultured for 1 day before compound treatment. PDO models were expanded for 1–5 passages, biobanked, embedded in hydrogel, and allowed to grow for 1 day; medium was refreshed on day 2 before treatment. Forty compounds were tested in each model at seven concentrations spanning 0.0000316–31.6 µM in 10-fold serial dilutions, with three replicate wells per concentration, followed by 5 days of drug exposure. Each screening plate included DMSO and/or PBS negative controls (8 wells per control type on each plate), a seven-point dose range of a standard-of-care reference compound, and 10 µM staurosporine as a positive cytotoxic control (8 wells per plate). Additional growth controls included serum-free conditions for the cell lines (4 wells per plate) and 0 ng/mL epithelial growth factor for PDOs (4 wells per plate). At the end of treatment, cell viability was measured using the CellTiter-Glo 3D ATP-release assay. ATP-release viability measurements of test conditions were normalized to those of negative controls on the same plate. The normalized viability values were used to calculate AUCs for the dose range of [100pM, 100µM]. AUC values were then normalized by the dose range to obtain normalized AUC values whose typical range was [0, 1].

\textbf{IMPROVE drug screening study 3}

Drug responses were evaluated in two colorectal cancer cell lines, nine colorectal cancer PDO lines (developed by the HCMI), and one sarcoma PDO line. The two cell lines, HT-29 and SW620, were seeded at 2,000 cells/well in 96-well plates using a 3:4 medium:Matrigel mixture and pre-cultured for 3 days in serum-free medium. The PDO screening was divided into two phases. In the first phase, five colorectal cancer PDO lines (PDM-2, PDM-5, PDM-7, PDM-8, and PDM-9) were expanded for approximately 3–4 passages and similarly seeded at 2,000 cells/well. In the second phase, four additional colorectal cancer PDO lines (PDM-4, PDM-95, PDM-103, and PDM-185) and one sarcoma PDO line (SARC0377TH) were expanded for approximately 1–2 passages. In drug screening experiments, colorectal cancer organoids were plated at 2,000 cells/well and sarcoma organoids at 5,000 cells/well. Both colorectal cancer cell lines and PDO lines were purchased from the ATCC.


For the cell line and the first PDO screens, 100 compounds were tested at seven concentrations from 100pM to 100µM in 10-fold serial dilutions, with three replicate wells per concentration. For the second organoid screen, compounds were tested at four concentrations from 0.01µM to 10µM, also in 10-fold dilutions and triplicate. Cells were treated twice on two consecutive days. Each plate contained 1\% DMSO negative controls (8 wells per plate) and 10 µM staurosporine positive controls (4 wells per plate). Twenty-four hours after the final treatment, viability was measured using the CellTiter-Glo 3D ATP-release assay. ATP-release viability measurements of test conditions were normalized to those of negative controls on the same plate. The normalized viability values were used to calculate AUCs for the dose range of [100pM, 100µM]. AUC values were then normalized by the dose range. Assay quality was evaluated using Z factor and robust Z factor. A threshold of 0.2 was specified for both Z factor metrics; plates were rerun if neither Z factor reached 0.2.

\textbf{Pilot 1 drug screening study} 

PDO lines were generated using an adapted epithelial cell-only submerged Matrigel culture method, as described in \cite{larsen2021pan}. Tumor specimens were transported in chilled RPMI and dissociated to single-cell suspensions using a GentleMACs instrument with the Miltenyi Tumor Dissociation Kit. Cells were embedded in growth-factor-reduced (GFR) Matrigel at 400 cells/µL and plated as 50-µL droplets in 24-well plates. After polymerization at 37°C, growth medium containing 10 µM Y-27632 was added. PDOs were maintained in chemically defined, serum-free Advanced DMEM/F12-based medium supplemented with B27, nicotinamide, N-acetylcysteine, recombinant growth factors including EGF, Noggin, R-spondin 1, Wnt-3A, and FGF-2, -7, and -10, and inhibitors of p38 MAPK and TGF\textbeta. Medium was changed twice weekly, and organoids were passaged every 1–2 weeks using TrypLE Express. Cultures showing sustained biomass expansion were cryopreserved for experiments after reaching $>10^6$ cells.

For drug screening, established PDO lines were dissociated into single cells and resuspended in a 30:70\% mixture of GFR Matrigel and growth medium. Approximately 2,000 cells were plated per well in 384-well plates and cultured for 72 hours before treatment. Drugs were prepared in growth medium containing 2.5 µM Caspase-3/7 green apoptosis assay reagent and added using an automated pipetting system. Seven 10-fold serial dilutions were prepared for each compound, with a maximum concentration of 100 µM. Organoids were then exposed to treatment for an additional 72 hours before high-content imaging.

After drug treatment, organoids were incubated with Hoechst 33342 and TO-PRO-3 for 1.5–2 hours and imaged at 10× magnification, capturing approximately 100–200 organoids per well. Z-stack images were projected into two dimensions and analyzed. Organoids and individual cells were identified from Hoechst staining, while dead/dying cells were identified using TO-PRO-3 and Caspase-3/7 signals. Viability was calculated as the percentage of viable cells within each organoid. Organoids containing fewer than three cells, the largest 1\% of organoids, and wells containing fewer than 20 organoids were excluded. Mean viability across retained organoids in a well was taken as the viability readout and normalized to vehicle controls on the same plate. The normalized viability values were used to calculate normalized AUCs for the dose range of [100 pM, 100 µM].

\subsection{2.1.2 Cell Line Drug Screening Data}

The cell line drug screening data were extracted from the IMPROVE benchmark \cite{Partin2026}. The benchmark integrates drug response data from five publicly available cell line drug screening studies: the Cancer Cell Line Encyclopedia (CCLE) \cite{Barretina2012, Ghandi2019}, the Cancer Therapeutics Response Portal version 2 (CTRPv2) \cite{Seashore-Ludlow2015, Rees2016}, the Genentech Cell Screening Initiative (gCSI) \cite{Klijn2015, Haverty2016}, and the Genomics of Drug Sensitivity in Cancer datasets GDSCv1 and GDSCv2 \cite{Iorio2016, Yang2013, Picco2019}. The numbers of experiments, drugs, and cell lines from each source are summarized in Table~\ref{tab:dataset_summary}. Drug responses are represented by the normalized AUC values calculated for the dose range of [100 pM, 100 µM]. 

\subsection{2.1.3 Omics Data and Drug Feature Data}

We performed RNA-seq experiments to generate gene expression data of PDO lines used in the four drug screening studies. TPM (transcripts per million reads mapped) values were first obtained and gene expression values were calculated as log$_2$(TPM + 1). Gene-level copy numbers, somatic mutations, and DNA methylations of HCMI PDO lines were retrieved from the Genomic Data Commons (GDC). Somatic mutations and gene-level copy numbers of PDO lines used in the Pilot 1 drug screening study were obtained using a targeted DNA-seq assay \cite{larsen2021pan}. Cell line omics data were extracted from the IMPROVE benchmark \cite{Partin2026}. The benchmark provides multiple omics modalities, including gene expressions, somatic mutations, copy numbers, and DNA methylations.

Drug feature data were also extracted from the the IMPROVE benchmark \cite{Partin2026}, including SMILES strings, binary molecular fingerprints, and numeric molecular descriptors. The 512-bit Extended Connectivity Fingerprints were generated by the Morgan algorithm using the open-source package RDKit \cite{RDKit}. The molecular descriptors were calculated using the open-source Mordred package \cite{Moriwaki2018}, and only 2D descriptors were kept for analysis. In our computation analysis, DRP models that require alternative drug representations, such as molecular graphs or SMILES-derived substructure sequences, generated these representations based on the drug SMILES strings. 

\subsection{2.2 Drug Response Prediction Models}


Our analysis utilized six publicly available DRP models: UNO \cite{xia2022uno}, IGTD \cite{zhu2021igtd}, GraphDRP \cite{Nguyen2021}, HiDRA \cite{Jin2021HiDRA}, DeepTTC \cite{jiang2022deeptta}, and LightGBM (LGBM) \cite{lgbm}. The models represent a range of machine learning and deep learning architectures, such as graph neural networks, convolutional neural networks, transformers, and gradient-boosted decision trees. All models use both cancer and compound features to predict drug response, with model-specific feature representations and architectures described in the Supplementary Information Section 1.
In the current study, gene expression was used as the cancer molecular representation, while drug representations included molecular descriptors, fingerprints, molecular graphs, or SMILES-derived substructure sequences depending on the model. Unless otherwise specified, models were trained using the mean squared error (MSE) loss with a batch size of 256 for up to 200 epochs, with early stopping applied using a patience of 20 epochs. LGBM was trained using a separate set of model-specific hyperparameters and early-stopping criteria. Additional model-specific preprocessing, architecture, and training details are provided in Supplementary Information Section 1. 

\subsection{2.3 Analysis Schemes}

\begin{figure}[ht]
    \centering
    \includegraphics[width=\textwidth]{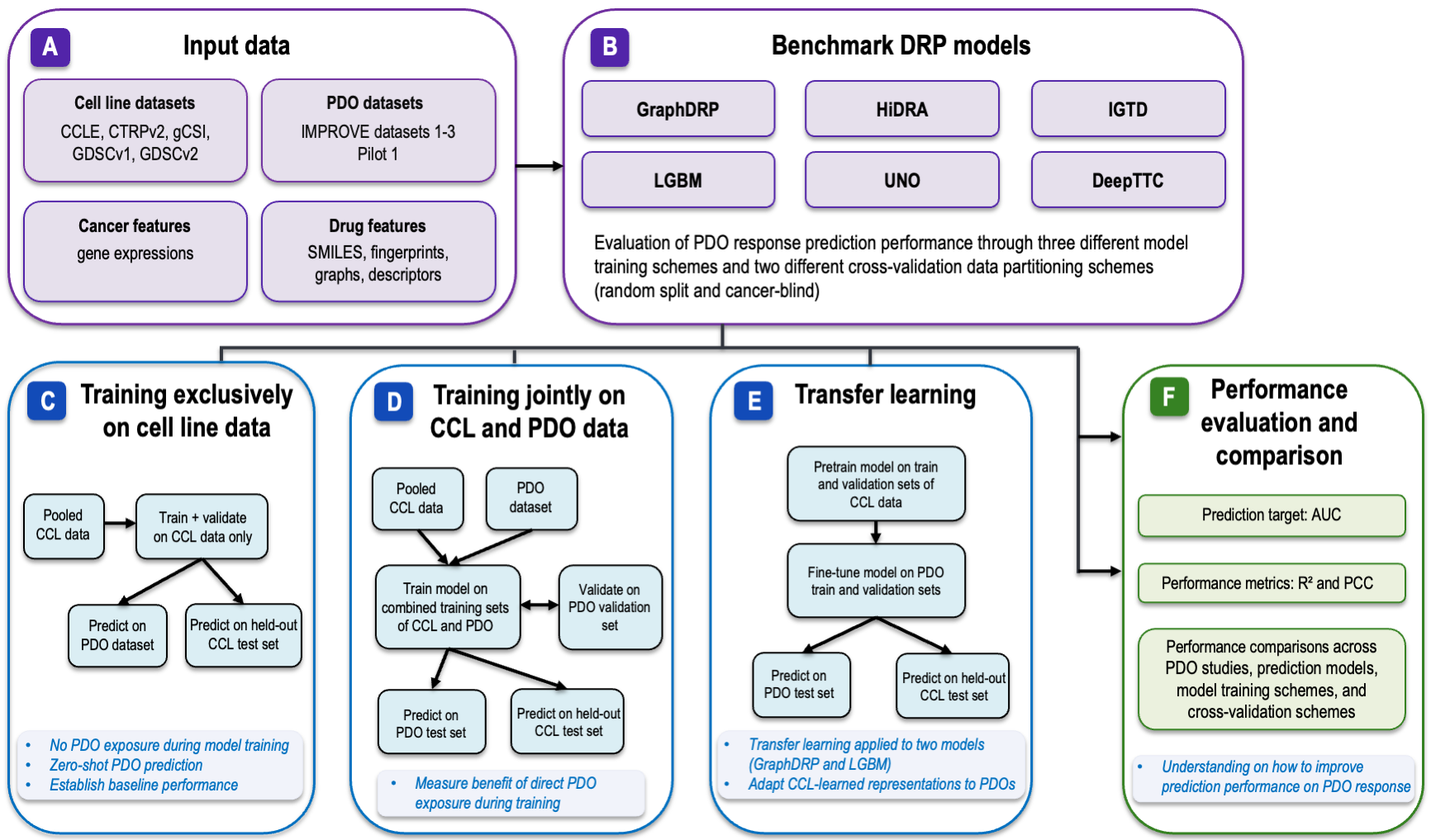}
    \caption{Overview of the DRP analysis framework. Six DRP models were used to predict PDO drug response using three learning strategies: cell line-only training, joint cell line and PDO training, and transfer learning with pretraining on cell lines followed by fine-tuning on PDOs. Random splits performed 10-fold cross-validations, whereas cancer-blind splits used $N$-fold data partitioning, where on IMPROVE datasets $N$ is the number of PDO lines in the dataset and on Pilot 1 dataset $N$ is 10.}
    \label{fig:workflow}
\end{figure}

We built DRP models and evaluated their prediction performance on PDO response under three different scenarios designed to assess how CCL and PDO data can be used for predicting PDO drug response: (1) training exclusively on CCL data, (2) joint training on CCL and PDO data, and (3) transfer learning with pretraining on CCL data followed by fine-tuning on PDO data (Figure~\ref{fig:workflow}). Two different data partition schemes, random split and cancer-blind split, were implemented for cross-validation analyses. In cancer-blind cross-validation, the train, validation and test sets did not share cancer cases, while this constrain did not apply in random cross-validation. 10-fold data partitioning with an 8:1:1 train, validation, and test ratio was used on CCL data in both random split and cancer-blind cross-validations. For PDO datasets, 10-fold data partitioning with an 8:1:1 train, validation, and test ratio was used in random cross-validation, while in cancer-blind cross-validation $N$-fold data partitioning with an $N-$2:1:1 train, validation, and test ratio was used, where $N$ is the number of organoid lines in the dataset. An exception was the Pilot 1 PDO dataset with 47 PDO lines, where 10-fold data partitioning with an 8:1:1 train, validation, and test ratio was also applied in the cancer-blind cross-validation. Predictive performance was quantified by comparing predicted and true normalized AUC values using the coefficient of determination ($R^2$) and the Pearson correlation coefficient (PCC). Prediction performances on PDO experiments and CCL experiments in the PDO drug screening studies were calculated separately. 

\subsubsection{2.3.1 Model Training on Cell Line Data Only}
\vspace{0.5em}

Since CCL drug screening data are available at a substantially greater scale than PDO data (See table~\ref{tab:dataset_summary}), we first evaluated the extent to which models trained exclusively on CCL data could generalize to PDO drug response prediction without target-domain adaptation. In this cell line only training scheme, no PDO molecular features and response data were used for model training. Each of the six DRP models was trained on the pooled CCL dataset comprising CCLE, CTRPv2, GDSCv1, GDSCv2, and gCSI data through cross-validation. Model performance was first evaluated on the CCL test set. The trained models were then applied to each PDO dataset to assess model generalizability to PDO response prediction. This analysis therefore provides a zero-shot evaluation of cross-domain generalization from CCLs to PDOs and serves as the baseline for assessing the benefit of incorporating PDO data through joint training (Section 2.3.2) or transfer learning (Section 2.3.3).

\subsubsection{2.3.2 Model Training on Cell Line and PDO Data Jointly}
\vspace{0.5em}

We next evaluated whether incorporating PDO data during model training could improve the prediction on PDO drug response. In the joint-training scheme, a separate model was trained by combining the train and validation sets of a CCL cross-validation partition with the train set of a PDO cross-validation partition, while the validation set of the PDO cross-validation partition was used for hyper-parameter optimization (HPO) and early stop of model training. The test set of the PDO cross-validation partition was used to evaluate the prediction performance on PDO response. The CCL experiments in PDO drug screening studies were never included in model training, but served as an additional test set for performance evaluation. As a quality control measure, we also evaluated the prediction performance on the test set in the CCL cross-validation partition. Under the random split scheme, the 10 CCL cross-validation partitions and the 10 PDO cross-validation partitions produced 100 models for each PDO dataset. Under the cancer-blind scheme, $10 \times N$ models were trained for each of the IMPROVE PDO datasets and 100 models were trained for the Pilot 1 PDO dataset.

\subsubsection{2.3.3 Transfer Learning from Cell Lines to PDOs}
\vspace{0.5em}

In the joint training scheme, the PDO data contribute to only a very small fraction of the train and validation sets used in model development, which might limit the benefit of including PDO data in the model training process. To address this limitation, we applied transfer learning, in which models were first pretrained on CCL data and then fine-tuned using PDO data. Specifically, each of the 10 models pretrained via the CCL cross-validation had its parameters updated through continuous training using the train and validation set of a PDO cross-validation partition. Then, the prediction performance of the updated model was evaluated using the test set of the PDO cross-validation partition. The CCL experiments in PDO drug screening studies were never included in model training, but served as an additional test set for performance evaluation. As a quality control measure, we also evaluated the prediction performance on the test set in the CCL cross-validation partition. Under the random split scheme, 100 fine-tuned models were generated for each PDO dataset. Under the cancer-blind scheme, $10 \times N$ fine-tuned models were trained for each of the IMPROVE PDO datasets and 100 models were trained for the Pilot 1 PDO dataset. 

Transfer learning analyses under both cross-validation schemes were conducted for GraphDRP and LGBM. For GraphDRP, we froze early layers in the model and updated the parameters of late layers during fine-tuning on PDO data \cite{NIPS2014_532a2f85}. The number of trainable layers was treated as a hyper-parameter to be optimized with other hyper-parameters using the differential evolution approach \cite{Storn1997DifferentialEvolution, Virtanen2020SciPy}. 
For LGBM, transfer learning was implemented by adding additional boosting steps to the pre-trained model. 
Additional details of the HPO process are provided in Supplementary Information Section 2.

\section{3  Results} 

\subsection{3.1 Correlation Between PDO Drug Screening Studies}

\begin{figure}[H]
    \centering
    \includegraphics[width=\textwidth]{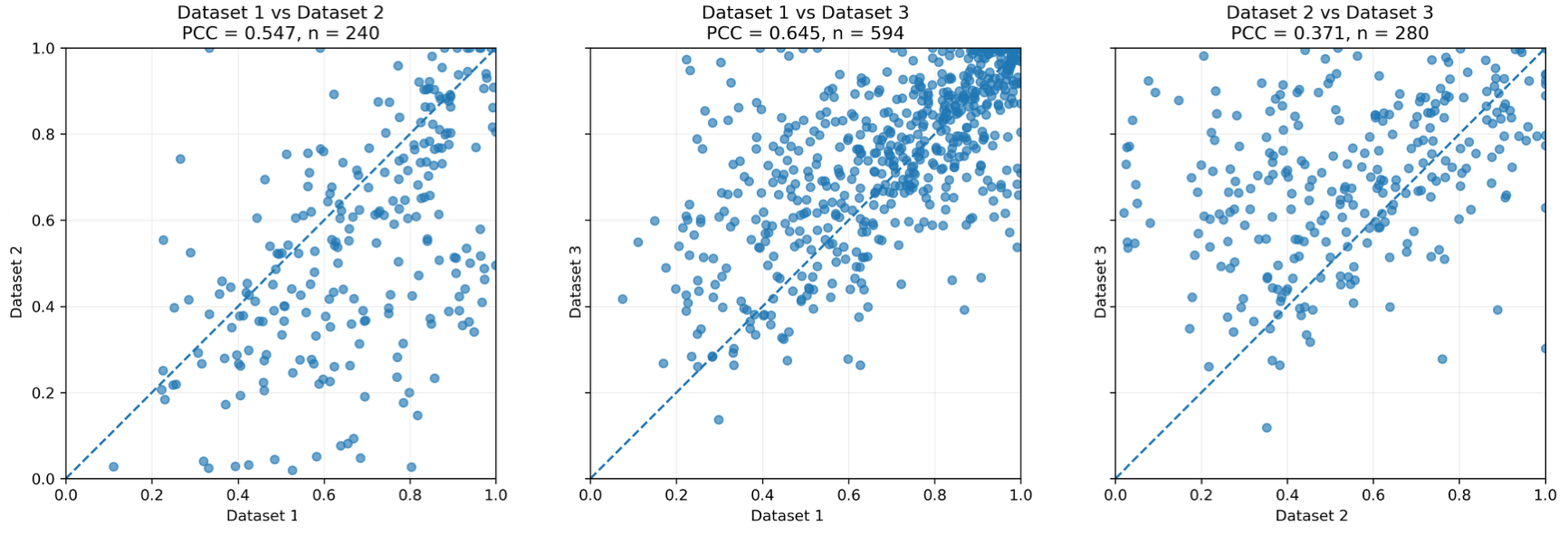}
    \caption{Pairwise comparison of drug response measurements across the three IMPROVE studies. Scatterplots show the normalized AUC values of matched experiments between a pair of IMPROVE datasets. Each point represents an experiment measured in both studies. PCC and the number of matched experiments are reported for each comparison.}
    \label{fig:source-correlation-heatmaps}
\end{figure}

As Figure~\ref{fig:improve_overlap} shows, the three IMPROVE drug screening studies share significant numbers of experiments, based on which we can evaluate the consistency of drug response measurements across studies. For each pair of the IMPROVE studies, we calculated the PCC of normalized AUC values based on experiments included in both studies, with a scatterplot visualization showing the distributions of AUC values (Figure~\ref{fig:source-correlation-heatmaps}). The correlation of response measurements varies across the three IMPROVE studies. The strongest correlation is observed between Datasets 1 and 3 (PCC = 0.645, n = 594), where n denotes the number of common experiments, followed by Datasets 1 and 2 (PCC = 0.547, n = 240). Datasets 2 and 3 exhibited the weakest correlation (PCC = 0.371, n = 280). All three correlations are at a moderate level, indicating a substantial variability remained between AUC measurements generated from studies using different experimental protocols and assay technologies.

\subsection{3.2 Prediction Performance}

\subsubsection{3.2.1 Model Training Using CCL Data Only}
\vspace{0.5em}

\begin{table}[!ht]
\centering
\caption{Prediction performance ($R^2$, mean $\pm$ s.d.) of models trained exclusively on CCL data in random split cross-validations. CCL Test indicates the test set of CCL data. Prediction performance on CCL experiments and PDO experiments in IMPROVE datasets 1-3 are listed separately.}
\label{tab:phase1_r2_random}
\resizebox{\linewidth}{!}{%
\begin{tabular}{lcccccccc}
\toprule
\multirow{2}{*}{\textbf{Model}} & \multirow{2}{*}{\textbf{CCL Test}} & \multicolumn{2}{c}{\textbf{IMPROVE Dataset 1}} & \multicolumn{2}{c}{\textbf{IMPROVE Dataset 2}} & \multicolumn{2}{c}{\textbf{IMPROVE Dataset 3}} & \multicolumn{1}{c}{\textbf{Pilot 1}} \\
\cmidrule(lr){3-4}\cmidrule(lr){5-6}\cmidrule(lr){7-8}\cmidrule(lr){9-9}
 & & CCL & PDO & CCL & PDO & CCL & PDO & PDO \\
\midrule
DeepTTC & 0.76$\,\pm\,$0.04 & 0.26$\,\pm\,$0.08 & 0.15$\,\pm\,$0.06 & -0.44$\,\pm\,$0.12 & -0.44$\,\pm\,$0.10 & -0.34$\,\pm\,$0.14 & 0.26$\,\pm\,$0.03 & 0.06$\,\pm\,$0.07 \\
UNO & 0.79$\,\pm\,$0.01 & 0.19$\,\pm\,$0.04 & 0.34$\,\pm\,$0.03 & -0.55$\,\pm\,$0.06 & -0.14$\,\pm\,$0.07 & -0.24$\,\pm\,$0.08 & 0.30$\,\pm\,$0.02 & 0.11$\,\pm\,$0.07 \\
GraphDRP & 0.78$\,\pm\,$0.00 & 0.26$\,\pm\,$0.03 & 0.31$\,\pm\,$0.03 & -0.56$\,\pm\,$0.05 & -0.07$\,\pm\,$0.06 & -0.18$\,\pm\,$0.09 & 0.20$\,\pm\,$0.04 & 0.03$\,\pm\,$0.06 \\
HiDRA & 0.77$\,\pm\,$0.00 & 0.28$\,\pm\,$0.08 & 0.03$\,\pm\,$0.18 & -0.58$\,\pm\,$0.12 & -0.65$\,\pm\,$0.32 & 0.04$\,\pm\,$0.15 & 0.12$\,\pm\,$0.13 & 0.08$\,\pm\,$0.10 \\
LGBM & 0.73$\,\pm\,$0.00 & 0.42$\,\pm\,$0.01 & 0.31$\,\pm\,$0.02 & -0.36$\,\pm\,$0.02 & -0.18$\,\pm\,$0.02 & 0.05$\,\pm\,$0.02 & 0.33$\,\pm\,$0.01 & 0.12$\,\pm\,$0.03 \\
IGTD & 0.78$\,\pm\,$0.00 & 0.24$\,\pm\,$0.04 & 0.33$\,\pm\,$0.05 & -0.57$\,\pm\,$0.06 & -0.09$\,\pm\,$0.05 & -0.17$\,\pm\,$0.09 & 0.22$\,\pm\,$0.06 & 0.02$\,\pm\,$0.08 \\
\midrule
\textit{Model Avg.} & 0.77$\,\pm\,$0.02 & 0.27$\,\pm\,$0.08 & 0.25$\,\pm\,$0.13 & -0.51$\,\pm\,$0.09 & -0.26$\,\pm\,$0.23 & -0.14$\,\pm\,$0.15 & 0.24$\,\pm\,$0.08 & 0.07$\,\pm\,$0.04 \\
\bottomrule
\end{tabular}%
}
\end{table}

\begin{table}[!ht]
\centering
\caption{Prediction performance ($R^2$, mean $\pm$ s.d.) of models trained exclusively on CCL data in cancer-blind cross-validations. CCL Test indicates the test set of CCL data. Prediction performance on CCL experiments and PDO experiments in IMPROVE datasets 1-3 are listed separately.}
\label{tab:phase1_r2_cancerblind}
\resizebox{\linewidth}{!}{%
\begin{tabular}{lcccccccc}
\toprule
\multirow{2}{*}{\textbf{Model}} & \multirow{2}{*}{\textbf{CCL Test}} & \multicolumn{2}{c}{\textbf{IMPROVE Dataset 1}} & \multicolumn{2}{c}{\textbf{IMPROVE Dataset 2}} & \multicolumn{2}{c}{\textbf{IMPROVE Dataset 3}} & \multicolumn{1}{c}{\textbf{Pilot 1}} \\
\cmidrule(lr){3-4}\cmidrule(lr){5-6}\cmidrule(lr){7-8}\cmidrule(lr){9-9}
 & & CCL & PDO & CCL & PDO & CCL & PDO & PDO \\
\midrule
DeepTTC & 0.64$\,\pm\,$0.01 & 0.40$\,\pm\,$0.02 & 0.24$\,\pm\,$0.06 & -0.06$\,\pm\,$0.04 & -0.29$\,\pm\,$0.09 & -0.44$\,\pm\,$0.08 & 0.30$\,\pm\,$0.04 & 0.12$\,\pm\,$0.09 \\
UNO & 0.69$\,\pm\,$0.01 & 0.38$\,\pm\,$0.04 & 0.25$\,\pm\,$0.06 & -0.19$\,\pm\,$0.05 & -0.22$\,\pm\,$0.11 & -0.17$\,\pm\,$0.08 & 0.31$\,\pm\,$0.04 & 0.11$\,\pm\,$0.06 \\
GraphDRP & 0.68$\,\pm\,$0.01 & 0.35$\,\pm\,$0.02 & 0.29$\,\pm\,$0.04 & -0.31$\,\pm\,$0.05 & -0.08$\,\pm\,$0.06 & -0.02$\,\pm\,$0.06 & 0.20$\,\pm\,$0.04 & -0.00$\,\pm\,$0.10 \\
HiDRA & 0.68$\,\pm\,$0.01 & 0.42$\,\pm\,$0.04 & 0.09$\,\pm\,$0.12 & -0.28$\,\pm\,$0.12 & -0.61$\,\pm\,$0.21 & -0.14$\,\pm\,$0.18 & 0.22$\,\pm\,$0.09 & 0.19$\,\pm\,$0.10 \\
LGBM & 0.70$\,\pm\,$0.01 & 0.44$\,\pm\,$0.01 & 0.30$\,\pm\,$0.03 & -0.27$\,\pm\,$0.04 & -0.19$\,\pm\,$0.05 & -0.08$\,\pm\,$0.05 & 0.31$\,\pm\,$0.01 & 0.10$\,\pm\,$0.04 \\
IGTD & 0.69$\,\pm\,$0.01 & 0.40$\,\pm\,$0.02 & 0.32$\,\pm\,$0.04 & -0.28$\,\pm\,$0.07 & -0.12$\,\pm\,$0.05 & -0.08$\,\pm\,$0.08 & 0.27$\,\pm\,$0.04 & 0.12$\,\pm\,$0.08 \\
\midrule
\textit{Model Avg.} & 0.68$\,\pm\,$0.02 & 0.40$\,\pm\,$0.03 & 0.25$\,\pm\,$0.08 & -0.23$\,\pm\,$0.09 & -0.25$\,\pm\,$0.19 & -0.16$\,\pm\,$0.15 & 0.27$\,\pm\,$0.05 & 0.11$\,\pm\,$0.06 \\
\bottomrule
\end{tabular}%
}
\end{table}

Table \ref{tab:phase1_r2_random} shows the prediction performance ($R^2$) in random split cross-validations. Models achieve an average $R^2$ ranging from 0.73 to 0.79 on the test set of CCL data with the deep learning models outperforming the conventional machine learning method LGBM. Performance decreases substantially when the same models are applied on data of PDO drug screening studies, including both CCL and PDO experiments in these studies. This demonstrates a challenge on model generalizability caused by both the biological gap between CCLs and PDOs and the difference of experimental protocols and assay technologies used in different drug screening studies. The average PDO prediction performance across models is similar on IMPROVE datasets 1 and 3 ($R^2 = 0.25$ and $0.24$), while the prediction performance on Pilot 1 dataset is lower ($R^2 = 0.07$) and the prediction performance on IMPROVE dataset 2 is the lowest ($R^2 = -0.26$). Performance difference across model architectures are more pronounced on PDO data than the test set of CCL data. The PDO prediction performance on IMPROVE dataset 1, IMPROVE dataset 3, and Pilot 1 has ranges of [0.03, 0.34], [0.12, 0.33], and [0.02, 0.12], respectively, while no model can always perform the best on all datasets.

In cancer-blind cross-validation (Table \ref{tab:phase1_r2_cancerblind}), prediction performance on the test set of CCL data is lower than that observed in random-split cross-validation, with an average $R^2$ across models of 0.68 compared with 0.77. This difference indicates that cancer-blind cross-validation, which simulates a precision oncology application, presents a more challenging prediction task than random-split cross-validation, which simulates a drug repurposing application. The prediction performance on PDO data measured by the model-average $R^2$ does not change much on all PDO datasets compared with the random-split cross-validation, which is reasonable because the PDO lines are not included in the CCL data for modeling training in both random split and cancer-blind cross-validations. In contrast, prediction performance on PDO data, measured by the average $R^2$ across models, does not change much compared with random-split cross-validation for all PDO datasets. This is expected because the PDO lines are not included in the CCL training data under either cross-validation scheme.

Prediction performances measured by PCC in the random split and cancer-blind cross-validations are reported in the Supplementary Information Section 3
and show the same general pattern as the $R^2$ prediction performance.

\subsubsection{3.2.2 Model Training Using CCL and PDO Data Jointly}
\vspace{0.5em}

\begin{table}[htbp]
\centering
\caption{Prediction performance ($R^2$, mean $\pm$ s.d.) of models trained using CCL and PDO data jointly in random split cross-validations. CCL Test indicates the test set of CCL data. Study CCL and Study PDO indicate the CCL experiments and the test set of PDO experiments in a PDO drug screening study, respectively.}
\label{tab:phase2_r2_random}
\resizebox{\linewidth}{!}{%
\begin{tabular}{lccccccccccc}
\toprule
\multirow{2}{*}{\textbf{Model}} & \multicolumn{3}{c}{\textbf{IMPROVE Dataset 1}} & \multicolumn{3}{c}{\textbf{IMPROVE Dataset 2}} & \multicolumn{3}{c}{\textbf{IMPROVE Dataset 3}} & \multicolumn{2}{c}{\textbf{Pilot 1}} \\
\cmidrule(lr){2-4}\cmidrule(lr){5-7}\cmidrule(lr){8-10}\cmidrule(lr){11-12}
& CCL Test & Study CCL & Study PDO & CCL Test & Study CCL & Study PDO & CCL Test & Study CCL & Study PDO & CCL Test & Study PDO \\
\midrule
DeepTTC & 0.78$\,\pm\,$0.01 & 0.39$\,\pm\,$0.06 & 0.74$\,\pm\,$0.11 & 0.76$\,\pm\,$0.02 & -0.51$\,\pm\,$0.10 & 0.55$\,\pm\,$0.20 & 0.78$\,\pm\,$0.01 & -0.03$\,\pm\,$0.10 & 0.73$\,\pm\,$0.04 & 0.77$\,\pm\,$0.01 & 0.74$\,\pm\,$0.06 \\
UNO & 0.79$\,\pm\,$0.01 & 0.30$\,\pm\,$0.04 & 0.77$\,\pm\,$0.09 & 0.77$\,\pm\,$0.02 & -0.50$\,\pm\,$0.08 & 0.58$\,\pm\,$0.23 & 0.79$\,\pm\,$0.01 & -0.11$\,\pm\,$0.10 & 0.74$\,\pm\,$0.04 & 0.79$\,\pm\,$0.01 & 0.77$\,\pm\,$0.04 \\
IGTD & 0.74$\,\pm\,$0.03 & 0.34$\,\pm\,$0.05 & 0.70$\,\pm\,$0.12 & 0.68$\,\pm\,$0.19 & -0.48$\,\pm\,$0.13 & 0.51$\,\pm\,$0.31 & 0.75$\,\pm\,$0.02 & 0.06$\,\pm\,$0.12 & 0.71$\,\pm\,$0.05 & 0.75$\,\pm\,$0.02 & 0.73$\,\pm\,$0.05 \\
HiDRA & 0.76$\,\pm\,$0.00 & 0.37$\,\pm\,$0.03 & 0.77$\,\pm\,$0.05 & 0.76$\,\pm\,$0.00 & -0.59$\,\pm\,$0.04 & 0.54$\,\pm\,$0.28 & 0.76$\,\pm\,$0.00 & 0.16$\,\pm\,$0.03 & 0.72$\,\pm\,$0.03 & 0.76$\,\pm\,$0.00 & 0.70$\,\pm\,$0.04 \\
LGBM & 0.74$\,\pm\,$0.01 & 0.45$\,\pm\,$0.01 & 0.69$\,\pm\,$0.02 & 0.74$\,\pm\,$0.00 & -0.37$\,\pm\,$0.01 & 0.58$\,\pm\,$0.02 & 0.74$\,\pm\,$0.01 & 0.16$\,\pm\,$0.01 & 0.69$\,\pm\,$0.04 & 0.74$\,\pm\,$0.00 & 0.72$\,\pm\,$0.00 \\
GraphDRP & 0.76$\,\pm\,$0.02 & 0.39$\,\pm\,$0.04 & 0.68$\,\pm\,$0.11 & 0.75$\,\pm\,$0.03 & -0.48$\,\pm\,$0.06 & 0.52$\,\pm\,$0.26 & 0.77$\,\pm\,$0.01 & -0.02$\,\pm\,$0.10 & 0.68$\,\pm\,$0.04 & 0.77$\,\pm\,$0.01 & 0.73$\,\pm\,$0.05 \\
\midrule
\textit{Model Avg.} & 0.76$\,\pm\,$0.02 & 0.37$\,\pm\,$0.05 & 0.72$\,\pm\,$0.04 & 0.74$\,\pm\,$0.03 & -0.49$\,\pm\,$0.07 & 0.55$\,\pm\,$0.03 & 0.77$\,\pm\,$0.02 & 0.04$\,\pm\,$0.11 & 0.71$\,\pm\,$0.02 & 0.76$\,\pm\,$0.02 & 0.73$\,\pm\,$0.02 \\
\bottomrule
\end{tabular}%
}
\end{table}

\begin{table}[htbp]
\centering
\caption{Prediction performance ($R^2$, mean $\pm$ s.d.) of models trained using CCL and PDO data jointly in cancer-blind cross-validations. CCL Test indicates the test set of CCL data. Study CCL and Study PDO indicate the CCL experiments and the test set of PDO experiments in a PDO drug screening study, respectively.}
\label{tab:phase2_r2_cancerblind}
\resizebox{\linewidth}{!}{%
\begin{tabular}{lccccccccccc}
\toprule
\multirow{2}{*}{\textbf{Model}} & \multicolumn{3}{c}{\textbf{IMPROVE Dataset 1}} & \multicolumn{3}{c}{\textbf{IMPROVE Dataset 2}} & \multicolumn{3}{c}{\textbf{IMPROVE Dataset 3}} & \multicolumn{2}{c}{\textbf{Pilot 1}} \\
\cmidrule(lr){2-4}\cmidrule(lr){5-7}\cmidrule(lr){8-10}\cmidrule(lr){11-12}
& CCL Test & Study CCL & Study PDO & CCL Test & Study CCL & Study PDO & CCL Test & Study CCL & Study PDO & CCL Test & Study PDO \\
\midrule
DeepTTC & 0.68$\,\pm\,$0.01 & 0.50$\,\pm\,$0.04 & 0.57$\,\pm\,$0.16 & 0.67$\,\pm\,$0.04 & -0.20$\,\pm\,$0.09 & -0.98$\,\pm\,$2.77 & 0.68$\,\pm\,$0.03 & -0.04$\,\pm\,$0.11 & 0.61$\,\pm\,$0.26 & 0.68$\,\pm\,$0.02 & 0.57$\,\pm\,$0.13 \\
UNO & 0.69$\,\pm\,$0.01 & 0.46$\,\pm\,$0.04 & 0.60$\,\pm\,$0.13 & 0.68$\,\pm\,$0.02 & -0.21$\,\pm\,$0.07 & -1.08$\,\pm\,$2.97 & 0.69$\,\pm\,$0.01 & -0.02$\,\pm\,$0.08 & 0.63$\,\pm\,$0.26 & 0.69$\,\pm\,$0.02 & 0.59$\,\pm\,$0.12 \\
IGTD & 0.31$\,\pm\,$0.54 & 0.42$\,\pm\,$0.12 & 0.40$\,\pm\,$0.11 & 0.34$\,\pm\,$0.55 & -0.16$\,\pm\,$0.16 & -1.48$\,\pm\,$3.74 & 0.64$\,\pm\,$0.04 & -0.01$\,\pm\,$0.17 & 0.56$\,\pm\,$0.31 & 0.64$\,\pm\,$0.03 & 0.56$\,\pm\,$0.10 \\
HiDRA & 0.67$\,\pm\,$0.01 & 0.50$\,\pm\,$0.02 & 0.46$\,\pm\,$0.14 & 0.67$\,\pm\,$0.01 & -0.37$\,\pm\,$0.05 & -1.82$\,\pm\,$4.55 & 0.68$\,\pm\,$0.01 & -0.03$\,\pm\,$0.07 & 0.60$\,\pm\,$0.20 & 0.68$\,\pm\,$0.01 & 0.54$\,\pm\,$0.08 \\
LGBM & 0.66$\,\pm\,$0.01 & 0.44$\,\pm\,$0.01 & 0.32$\,\pm\,$0.02 & 0.69$\,\pm\,$0.01 & -0.31$\,\pm\,$0.03 & -1.61$\,\pm\,$0.32 & 0.62$\,\pm\,$0.01 & 0.07$\,\pm\,$0.02 & 0.54$\,\pm\,$0.01 & 0.68$\,\pm\,$0.02 & 0.61$\,\pm\,$0.02 \\
GraphDRP & 0.68$\,\pm\,$0.01 & 0.45$\,\pm\,$0.04 & 0.51$\,\pm\,$0.15 & 0.66$\,\pm\,$0.04 & -0.29$\,\pm\,$0.07 & -1.23$\,\pm\,$3.45 & 0.68$\,\pm\,$0.01 & 0.02$\,\pm\,$0.11 & 0.59$\,\pm\,$0.25 & 0.68$\,\pm\,$0.01 & 0.57$\,\pm\,$0.13 \\
\midrule
\textit{Model Avg.} & 0.61$\,\pm\,$0.15 & 0.46$\,\pm\,$0.03 & 0.48$\,\pm\,$0.11 & 0.62$\,\pm\,$0.14 & -0.26$\,\pm\,$0.08 & -1.37$\,\pm\,$0.33 & 0.66$\,\pm\,$0.03 & 0.00$\,\pm\,$0.04 & 0.59$\,\pm\,$0.03 & 0.67$\,\pm\,$0.02 & 0.57$\,\pm\,$0.02 \\
\bottomrule
\end{tabular}%
}
\end{table}

Tables~\ref{tab:phase2_r2_random} shows the $R^2$ prediction performance of models trained using CCL and PDO data jointly in random split cross-validations. Compared with training models using only CCL data (Table \ref{tab:phase1_r2_random}), incorporating PDO data directly into model training substantially improves PDO prediction performance. The model-average $R^2$ for PDO prediction increases from $0.25$ to $0.72$ for IMPROVE dataset 1, from $-0.26$ to $0.55$ for IMPROVE dataset 2, from $0.24$ to $0.71$ for IMPROVE dataset 3, and from $0.07$ to $0.73$ for Pilot 1 dataset. Prediction performance on the test set of CCL data averaged across models and studies is $R^2 = 0.7575$, which is similar to that of using only CCL data to train models ($R^2 = 0.77$), indicating that the improvement on PDO prediction does not require a substantial loss of performance in predicting CCL response. Prediction performance on the CCL experiments from the IMPROVE PDO drug screening studies also improves slightly when PDO data are included in model training, with $R^2$ increasing from $0.27$ to $0.37$ for IMPROVE Dataset 1, from $-0.51$ to $-0.49$ for IMPROVE Dataset 2, and from $-0.14$ to $0.04$ for IMPROVE Dataset 3. This result suggests that incorporating PDO experiments conducted using the same experimental protocols and assay technologies as the CCL experiments into model training can improve model generalizability to these CCL experiments.

The $R^2$ prediction performance in the cancer-blind cross-validation largely follows the same pattern of the prediction performance obtained in the random split cross-validation. When PDO data are included in model training, PDO  prediction performance increases substantially on three out of the four PDO drug screening datasets, increasing from $0.25$ to $0.48$ on IMPROVE dataset 1, from $0.27$ to $0.59$ on IMPROVE dataset 2, and from $0.11$ to $0.57$ on Pilot 1, measured by average $R^2$ across models. However, the PDO prediction performance in cancer-blind cross-validation is lower than that of random split cross-validation, which again indicates model generalizability across cancer cases is more challenging. 

Across model architectures, UNO always achieves the highest or tied-highest PDO prediction performance on all four datasets in random split corss-validation and two out of the four datasets in cancer-blind cross-validation. Overall, joint training reduces the CCL-to-PDO prediction gap for most datasets, but its effectiveness depends on whether predictions involve previously observed or entirely unseen PDO lines. Prediction performance measured by PCC generally supports the same trends observed on $R^2$ prediction performance results and is provided in the Supplementary Information Section 3.

\subsubsection{3.2.3 Transfer Learning From CCLs To PDOs}
\vspace{0.5em}

\begin{table}[htbp]
\centering
\caption{Transfer learning prediction performance ($R^2$, mean $\pm$ s.d.) of models in random split cross-validations. CCL Test indicates the test set of CCL data. Study CCL and Study PDO indicate the CCL experiments and the test set of PDO experiments in a PDO drug screening study, respectively. Pilot 1 study does not include Study CCL experiments.}
\label{tab:phase3_r2_random}
\resizebox{\linewidth}{!}{%
\begin{tabular}{lccccccccccc}
\toprule
\multirow{2}{*}{\textbf{Model}}
& \multicolumn{3}{c}{\textbf{IMPROVE Dataset 1}}
& \multicolumn{3}{c}{\textbf{IMPROVE Dataset 2}}
& \multicolumn{3}{c}{\textbf{IMPROVE Dataset 3}}
& \multicolumn{2}{c}{\textbf{Pilot 1}} \\
\cmidrule(lr){2-4}
\cmidrule(lr){5-7}
\cmidrule(lr){8-10}
\cmidrule(lr){11-12}
& CCL Test & Study CCL & Study PDO
& CCL Test & Study CCL & Study PDO
& CCL Test & Study CCL & Study PDO
& CCL Test & Study PDO \\
\midrule
GraphDRP
& -0.13$\,\pm\,$1.25 & 0.56$\,\pm\,$0.06 & 0.79$\,\pm\,$0.06
& 0.49$\,\pm\,$0.15 & -0.05$\,\pm\,$0.15 & 0.55$\,\pm\,$0.28
& 0.56$\,\pm\,$0.08 & 0.14$\,\pm\,$0.13 & 0.73$\,\pm\,$0.05
& 0.47$\,\pm\,$0.07 & 0.76$\,\pm\,$0.04 \\

LGBM
& -1.16$\,\pm\,$0.53 & 0.65$\,\pm\,$0.02 & 0.75$\,\pm\,$0.08
& 0.10$\,\pm\,$0.07 & 0.16$\,\pm\,$0.17 & 0.64$\,\pm\,$0.16
& 0.44$\,\pm\,$0.07 & -0.12$\,\pm\,$0.08 & 0.73$\,\pm\,$0.05
& 0.07$\,\pm\,$0.16 & 0.75$\,\pm\,$0.04 \\
\bottomrule
\end{tabular}%
}
\end{table}

\begin{table}[htbp]
\centering
\caption{Transfer learning prediction performance ($R^2$, mean $\pm$ s.d.) of models in cancer-blind cross-validations. CCL Test indicates the test set of CCL data. Study CCL and Study PDO indicate the CCL experiments and the test set of PDO experiments in a PDO drug screening study, respectively. Pilot 1 study does not include Study CCL experiments.}
\label{tab:phase3_r2_cancerblind}
\resizebox{\linewidth}{!}{%
\begin{tabular}{lccccccccccc}
\toprule
\multirow{2}{*}{\textbf{Model}}
& \multicolumn{3}{c}{\textbf{IMPROVE Dataset 1}}
& \multicolumn{3}{c}{\textbf{IMPROVE Dataset 2}}
& \multicolumn{3}{c}{\textbf{IMPROVE Dataset 3}}
& \multicolumn{2}{c}{\textbf{Pilot 1}} \\
\cmidrule(lr){2-4}
\cmidrule(lr){5-7}
\cmidrule(lr){8-10}
\cmidrule(lr){11-12}
& CCL Test & Study CCL & Study PDO
& CCL Test & Study CCL & Study PDO
& CCL Test & Study CCL & Study PDO
& CCL Test & Study PDO \\
\midrule
GraphDRP
& 0.04$\,\pm\,$0.39 & 0.59$\,\pm\,$0.07 & 0.58$\,\pm\,$0.17
& 0.32$\,\pm\,$0.18 & 0.06$\,\pm\,$0.12 & -1.20$\,\pm\,$3.37
& 0.38$\,\pm\,$0.13 & 0.14$\,\pm\,$0.12 & 0.60$\,\pm\,$0.23
& 0.43$\,\pm\,$0.14 & 0.57$\,\pm\,$0.12 \\

LGBM
& -0.37$\,\pm\,$0.75 & 0.61$\,\pm\,$0.07 & 0.50$\,\pm\,$0.22
& 0.28$\,\pm\,$0.29 & 0.17$\,\pm\,$0.17 & -1.44$\,\pm\,$3.69
& 0.38$\,\pm\,$0.12 & -0.27$\,\pm\,$0.20 & 0.60$\,\pm\,$0.32
& 0.05$\,\pm\,$0.18 & 0.63$\,\pm\,$0.09 \\
\bottomrule
\end{tabular}%
}
\end{table}

Tables~\ref{tab:phase3_r2_random} shows the transfer learning prediction performance ($R^2$) in random split cross-validations. Compared with joint model training using CCL and PDO data (Tables~\ref{tab:phase2_r2_random}), transfer learning further improves the PDO prediction performance for both GraphDRP and LGBM models on all four PDO datasets with an improvement range of [0.03, 0.11]. This result indicates that fine-tuning models on PDO data through transfer learning is more effective in improving PDO prediction performance than training models on combined PDO and CCL data, in which PDO data constitute only a very small fraction of the training and validation data. For the prediction performance on CCL experiments from PDO drug screening studies, transfer learning improves the performance of GraphDRP on all three IMPROVE studies and improve the performance of LGBM on two of the IMPROVE studies. The prediction performance on the CCL test set decreases substantially for both GraphDRP and LGBM across all four PDO studies compared with joint training on CCL and PDO data, indicating that the improvement in PDO prediction performance achieved through transfer learning comes at the cost of forgetting previously learned CCL drug response patterns.

Tables~\ref{tab:phase3_r2_cancerblind} shows the transfer learning prediction performance ($R^2$) in cancer-blind cross-validations. Compared with model training using combined CCL and PDO data (Tables~\ref{tab:phase2_r2_cancerblind}), transfer learning improves the PDO prediction performance of GraphDRP on three out of the four PDO datasets by a range of [0.01, 0.07] and improves the PDO prediction performance of LGBM on all four PDO datasets by a range of [0.02, 0.18]. As observed in random-split cross-validation, transfer learning also improves prediction performance in cancer-blind cross-validation on the CCL experiments from the PDO drug screening studies, with improvements observed for GraphDRP across all three IMPROVE studies and for LGBM in two of the three studies. Similar to the results observed in random-split cross-validation, prediction performance on the CCL test sets decreases substantially compared with joint training on CCL and PDO data, further highlighting the trade-off between CCL and PDO prediction performance in transfer learning.

Performance results measured by PCC are generally consistent with the patterns observed in the $R^2$ results and are reported in the Supplementary Information Section 3.

\section{4 Discussion}

A major contribution of our study is the generation and evaluation of new PDO drug screening datasets for DRP research. Large-scale CCL pharmacogenomic datasets have been widely used for developing DRP models, but they do not fully capture the biological context in which these models are ultimately intended to be applied. PDOs provide an experimental system that more closely preserves the biological characteristics of patient tumors. However, PDO drug screening data remain limited, and their utility for drug response modeling has not been extensively investigated. We conducted four PDO drug screening studies using multiple screening platforms and experimental settings, generating drug response measurements along with molecular profiles of the cancer models and drug feature data. These studies focused on colorectal cancer and head and neck cancer, for which existing DRP models have shown limited prediction accuracy. The compounds used in the screening studies were selected to cover major MoAs relevant to current cancer therapies and oncology research.

We evaluated the utility of our datasets for developing and evaluating DRP models for PDO response prediction using three training strategies: (1) training models exclusively on CCL data, (2) training models jointly on CCL and PDO data, and (3) pretraining models on CCL data followed by fine-tuning on PDO data. Models trained exclusively on CCL data generally retained limited predictive power on PDO datasets, indicating that some molecular and pharmacological relationships learned from CCL screening data can transfer to PDO systems. Incorporating a portion of the PDO data into model training improved PDO response prediction performance, demonstrating that even though the target-domain data constituted only a small fraction of the overall training set, they provided valuable information that was not captured by CCL screening data but was important for accurate PDO response prediction. Transfer learning, in which models were pretrained on CCL data and subsequently fine-tuned on PDO data, generally further improved PDO response prediction performance, although the magnitude of improvement varied across models and datasets.

We evaluated the models under two different data partitioning schemes: random-split cross-validation and cancer-blind cross-validation. In general, prediction performance was higher under random-split cross-validation than under cancer-blind cross-validation when all other factors, such as model, training data, and training strategy, were held constant. This finding is expected, because generalizing to unseen cancer cases is more challenging than predicting responses for cancer cases represented in the training data. The patterns of variation in prediction performance across models, datasets, and training strategies are largely consistent between random-split and cancer-blind cross-validation.

PDO prediction performance on IMPROVE Dataset 2 was consistently the lowest among the four drug screening studies, with negative $R^2$ values observed in some cases, indicating that the models performed worse than a predictor that always predicts the mean response. Upon reviewing the experimental procedures, we identified an issue with insufficient mixing of the cell culture medium, which resulted in uneven cell seeding across wells and introduced substantial noise into the data. We believe this issue was a major contributor to the poor prediction performance on this dataset. Interestingly, even when $R^2$ values were low or negative, PCC prediction performance sometimes remained moderate to high, suggesting that the models could capture the relative ranking of drug responses despite failing to accurately predict their magnitudes.

Our IMPROVE drug screening studies include overlapping experiments across studies using different screening platforms and experimental settings, allowing an evaluation on the reproducibility of drug response measurements. Moderate levels of correlations in the range of [0.371, 0.645] are observed on response measurements from different studies, highlighting an important challenge for both drug screening experiments and PDO response modeling. Reproducibility of drug screening experiments need to be improved. Future efforts to expand PDO pharmacogenomic resources should consider experimental standardization, consistent quality-control procedure, and study or site effects as integral components of benchmark design. Most computational DRP models can only learn relationships that are represented consistently in their target data, and variations of response measurements caused by differences in screening platforms may therefore limit model generalizability even when the underlying biological response is similar. AI approaches that explicitly consider measurement variations may be useful for DRP tasks across datasets. Nevertheless, datasets generated across heterogeneous experimental settings remain valuable because they provide a more stringent test of whether predictive relationships are robust to realistic sources of measurement variation.

We have observed that transfer learning can improve prediction performance in the target PDO domain compared with both training exclusively on CCL data and joint training on CCL and PDO data, but at the cost of forgetting previously learned CCL response patterns. More sophisticated transfer learning approaches may improve prediction performance in the target domain while preserving performance in the source domain. One such approach is source replay, which incorporates data from both the source and target domains during model fine-tuning, with target-domain data accounting for a substantial proportion of the fine-tuning dataset. Because source-domain datasets are typically much larger than target-domain datasets, a small subset of source-domain data can be randomly sampled and combined with target-domain data for fine-tuning.

Based on the findings of our study, several directions for future research can be identified. First, new computational methods are needed to improve cross-domain and cross-study generalization, particularly by accounting for systematic differences in experimental platforms and drug response distributions. Second, although our current DRP analyses use only gene expression data to characterize cancer models, we have made additional multi-omics data for CCLs and PDOs, including copy number alterations, somatic mutations, and DNA methylation profiles, available through our online data repository. Incorporating these additional molecular modalities into DRP modeling may improve prediction performance and generalizability. Third, despite our efforts to generate new PDO drug screening and molecular datasets, available data resources for PDO drug response modeling remain limited. Substantially larger and more diverse datasets spanning a broader range of cancer types, PDO models, and compounds are needed to support the development of accurate and generalizable DRP models. Fourth, further research is needed to bridge the gap between PDO drug response modeling and the prediction of treatment responses in patients. Integrating CCL, PDO, and patient-level data for predicting drug responses in patient tumors will require more sophisticated computational approaches, such as multi-domain transfer learning.

Overall, our work provides both new datasets and a computational framework for PDO drug response modeling. Our results demonstrate that large-scale CCL datasets remain valuable for learning generalizable predictive relationships but are insufficient to ensure robust prediction of PDO responses. Incorporating even limited PDO data into model training can substantially improve prediction performance in the target domain. At the same time, variability in drug response measurements arising from differences in experimental protocols and technologies, together with reduced generalization performance on unseen cancer cases, highlights two distinct challenges that must be addressed as DRP models advance toward applications in therapeutic development and patient care. Expanding and harmonizing PDO drug screening datasets, while developing computational methods that explicitly account for these sources of variability, will be important steps toward building DRP models that more effectively leverage patient-derived experimental data.

\section{Data and Code Availability}

\begin{enumerate}
    \item \textbf{Code.} The code used for training and evaluating DRP models in this paper is available here: \url{https://github.com/zhuyitan/PDO_DRP_Analysis} 
    
    \item \textbf{Data.} The full dataset is available at: \url{https://doi.org/10.5281/zenodo.22920452}
\end{enumerate}

\section{Acknowledgement}

This research was part of the Innovative Methodologies and New Data for Predictive Oncology Model Evaluation (IMPROVE) project under the National Cancer Institute - Department of Energy (NCI-DOE) collaboration program that has been funded in whole or in part with federal funds from the National Cancer Institute, National Institutes of Health, Task Order No. 75N91019F00134 and from the Department of Energy under Award Number ACO22002-001-00000.

\printbibliography

@article{Partin2026,
  author = {Partin, Alexander and Vasanthakumari, Priyanka and others},
  title = {Benchmarking community drug response prediction models: datasets, models, tools, and metrics for cross-dataset generalization analysis},
  journal = {Briefings in Bioinformatics},
  volume = {27},
  number = {1},
  year = {2026},
  doi = {10.1093/bib/bbaf667}
}

@article{Partin2023,
  author = {Partin, A and Brettin, T S and Zhu, Y and Narykov, O and others},
  title = {Deep learning methods for drug response prediction in cancer: Predominant and emerging trends},
  journal = {Frontiers in Medicine},
  volume = {10},
  year = {2023}
}

@article{Nguyen2021,
  author = {Nguyen, T. and Nguyen, G. T. T. and others},
  title = {Graph Convolutional Networks for Drug Response Prediction},
  journal = {IEEE/ACM Transactions on Computational Biology and Bioinformatics},
  year = {2021}
}

@article{Barretina2012,
  author = {Barretina, Jordi and Caponigro, Giordano and Stransky, Nicolas and Venkatesan, Kavitha and Margolin, Adam and Kim, Sungjoon and Wilson, Christopher and Leh{\'a}r, Joseph and Kryukov, Gregory and Sonkin, Dmitriy and others},
  title = {The Cancer Cell Line Encyclopedia Enables Predictive Modelling of Anticancer Drug Sensitivity},
  journal = {Nature},
  volume = {483},
  number = {7391},
  pages = {603--607},
  year = {2012}
}

@article{Ghandi2019,
  author = {Ghandi, Mahmoud and Huang, Franklin W and Jane-Valbuena, Judit and Kryukov, Gregory V and others},
  title = {Next-generation characterization of the Cancer Cell Line Encyclopedia},
  journal = {Nature},
  volume = {569},
  pages = {503--508},
  year = {2019}
}

@article{Iorio2016,
  author = {Iorio, F and Knijnenburg, T A and Vis, D J and Bignell, G R and Menden, M P and Schubert, M and others},
  title = {A Landscape of Pharmacogenomic Interactions in Cancer},
  journal = {Cell},
  year = {2016}
}

@article{Yang2013,
  author = {Yang, W and Soares, J and Greninger, P and Edelman, E J and Lightfoot, H and Forbes, S and others},
  title = {Genomics of Drug Sensitivity in Cancer (GDSC): a resource for therapeutic biomarker discovery in cancer cells},
  journal = {Nucleic Acids Research},
  year = {2013}
}

@article{Picco2019,
  author = {Picco, G and Chen, E D and Alonso, L G and Behan, F M and Gon{\c{c}}alves, E and Bignell, G and others},
  title = {Functional linkage of gene fusions to cancer cell fitness assessed by pharmacological and CRISPR-Cas9 screening},
  journal = {Nature Communications},
  year = {2019}
}

@article{Moriwaki2018,
  author = {Moriwaki, H and Tian, Y-S and others},
  title = {Mordred: A molecular descriptor calculator},
  journal = {Journal of Cheminformatics},
  volume = {10},
  year = {2018}
}

@article{Klijn2015,
  author = {Klijn, C and Durinck, S and Stawiski, E W and Haverty, P M and Jiang, Z and Liu, H and others},
  title = {A comprehensive transcriptional portrait of human cancer cell lines},
  journal = {Nature Biotechnology},
  volume = {33},
  number = {3},
  pages = {306--312},
  year = {2015}
}

@article{Haverty2016,
  author = {Haverty, P M and Lin, E and Tan, J and Yu, Y and Lam, B and Lianoglou, S and others},
  title = {Reproducible pharmacogenomic profiling of cancer cell line panels},
  journal = {Nature},
  volume = {533},
  pages = {333--337},
  year = {2016}
}

@article{Seashore-Ludlow2015,
  author = {Seashore-Ludlow, B and Rees, M G and Cheah, J H and Cokol, M and Price, E V and Coletti, M E and others},
  title = {Harnessing Connectivity in a Large-Scale Small-Molecule Sensitivity Dataset},
  journal = {Cancer Discovery},
  volume = {5},
  number = {11},
  pages = {1210--1223},
  year = {2015}
}

@article{Rees2016,
  author = {Rees, M G and Seashore-Ludlow, B and Cheah, J H and Adams, D J and Price, E V and Gill, S and others},
  title = {Correlating chemical sensitivity and basal gene expression reveals mechanism of action},
  journal = {Nature Chemical Biology},
  volume = {12},
  pages = {109--116},
  year = {2016}
}

@article{Weil2024,
  author = {Weil, M and Partin, A and Kousa, N and others},
  title = {The IMPROVE framework for automating and standardizing tumor drug response prediction modeling},
  journal = {European Journal of Cancer},
  volume = {211},
  year = {2024}
}

@misc{RDKit,
  title = {RDKit: Open-source cheminformatics},
  howpublished = {\url{https://www.rdkit.org}}
}

@article{xia2022uno,
  title={A cross-study analysis of drug response prediction in cancer cell lines},
  author={Xia, Fangfang and Allen, Jonathan and Balaprakash, Prasanna and Brettin, Thomas and Garcia-Cardona, Cristina and Clyde, Austin and Cohn, Judith and Doroshow, James and Duan, Xiaotian and Dubinkina, Veronika and Evrard, Yvonne and Fan, Yun Jae and Gans, Jason and He, Stewart and Lu, Pinyi and Maslov, Sergei and Partin, Alexander and Shukla, Maulik and Stahlberg, Eric and Wozniak, Justin M. and Yoo, Hyunseung and Zaki, George and Zhu, Yitan and Stevens, Rick},
  journal={Briefings in Bioinformatics},
  volume={23},
  number={1},
  pages={bbab356},
  year={2022},
  publisher={Oxford University Press},
  doi={10.1093/bib/bbab356}
}

@article{zhu2021igtd,
  title={Converting tabular data into images for deep learning with convolutional neural networks},
  author={Zhu, Yitan and Brettin, Thomas and Xia, Fangfang and Partin, Alexander and Shukla, Maulik and Yoo, Hyunseung and Evrard, Yvonne A. and Doroshow, James H. and Stevens, Rick L.},
  journal={Scientific Reports},
  volume={11},
  pages={11325},
  year={2021},
  publisher={Nature Publishing Group},
  doi={10.1038/s41598-021-90923-y}
}

@article{jiang2022deeptta,
  title={DeepTTA: a transformer-based model for predicting cancer drug response},
  author={Jiang, Likun and Jiang, Changzhi and Yu, Xinyu and Fu, Rao and Jin, Shuting and Liu, Xiangrong},
  journal={Briefings in Bioinformatics},
  volume={23},
  number={3},
  pages={bbac100},
  year={2022},
  publisher={Oxford University Press},
  doi={10.1093/bib/bbac100}
}

@article{subramanian2017lincs,
  title={A Next Generation Connectivity Map: L1000 Platform and the First 1,000,000 Profiles},
  author={Subramanian, Aravind and Narayan, Rajiv and Corsello, Steven M. and Peck, David D. and Natoli, Ted E. and Lu, Xiaodong and Gould, Joshua and Davis, John F. and Tubelli, Andrew A. and Asiedu, Jacob K. and others},
  journal={Cell},
  volume={171},
  number={6},
  pages={1437--1452.e17},
  year={2017},
  publisher={Elsevier},
  doi={10.1016/j.cell.2017.10.049}
}

@article{larsen2021pan,
  title={A pan-cancer organoid platform for precision medicine},
  author={Larsen, Brian M and Kannan, Madhavi and Langer, Lee F and Leibowitz, Benjamin D and Bentaieb, Aicha and Cancino, Andrea and Dolgalev, Igor and Drummond, Bridgette E and Dry, Jonathan R and Ho, Chi-Sing and others},
  journal={Cell reports},
  volume={36},
  number={4},
  year={2021},
  publisher={Elsevier}
}

@Manual{lgbm,
  title = {lightgbm: Light Gradient Boosting Machine},
  author = {Yu Shi and Guolin Ke and Damien Soukhavong and James Lamb and Qi Meng and Thomas Finley and Taifeng Wang and Wei Chen and Weidong Ma and Qiwei Ye and Tie-Yan Liu and Nikita Titov and David Cortes},
  year = {2026},
  note = {R package version 4.7.0.99},
  url = {https://github.com/lightgbm-org/LightGBM},
}

@article{BenDavid2018,
  author  = {Ben-David, U. and Siranosian, B. and Ha, G. and others},
  title   = {Genetic and transcriptional evolution alters cancer cell line drug response},
  journal = {Nature},
  volume  = {560},
  pages   = {325--330},
  year    = {2018},
  doi     = {10.1038/s41586-018-0409-3}
}

@article{Jin2021HiDRA,
  author  = {Jin, I and Nam, H},
  title   = {HiDRA: hierarchical network for drug response prediction with attention},
  journal = {J Chem Inf Model},
  year    = {2021},
  volume  = {61},
  pages   = {3858--3867},
  doi     = {10.1021/acs.jcim.1c00706}
}

@article{Storn1997DifferentialEvolution,
  author  = {Storn, R. and Price, K.},
  title   = {Differential Evolution -- A Simple and Efficient Heuristic for Global Optimization over Continuous Spaces},
  journal = {Journal of Global Optimization},
  year    = {1997},
  volume  = {11},
  pages   = {341--359},
  doi     = {10.1023/A:1008202821328}
}

@article{Virtanen2020SciPy,
  author  = {Virtanen, Pauli and Gommers, Ralf and Oliphant, Travis E. and others},
  title   = {SciPy 1.0: Fundamental Algorithms for Scientific Computing in Python},
  journal = {Nature Methods},
  year    = {2020},
  volume  = {17},
  number  = {3},
  pages   = {261--272},
  doi     = {10.1038/s41592-019-0686-2}
}

@inproceedings{NIPS2014_532a2f85,
 author = {Yosinski, Jason and Clune, Jeff and Bengio, Yoshua and Lipson, Hod},
 booktitle = {Advances in Neural Information Processing Systems},
 editor = {Z. Ghahramani and M. Welling and C. Cortes and N. Lawrence and K. Weinberger},
 pages = {},
 publisher = {Curran Associates, Inc.},
 title = {How transferable are features in deep neural networks?},
 url = {https://proceedings.neurips.cc/paper_files/paper/2014/file/532a2f85b6977104bc93f8580abbb330-Paper.pdf},
 volume = {27},
 year = {2014}
}

\clearpage
\section{Supplementary Information}
\label{sec:supp-info}

\subsection{1 Details of Drug Response Prediction Models}
\label{sec:supp-models}

Six publicly available DRP models were evaluated using implementations within the IMPROVE framework \cite{Partin2026}. The models differ in their neural network architectures and in how they represent cancers and compounds. Unless otherwise specified, deep learning models were optimized using the mean squared error (MSE) loss, a batch size of 256 when training on CCL data, and a maximum of 200 training epochs, with early stopping applied after 20 epochs without a significant improvement on validation performance.

\subsubsection{1.1 UNO}

UNO \cite{xia2022uno} is a feed-forward deep neural network consisting of separate fully-connected subnetworks for cancer cases and drugs. The outputs of these subnetworks are concatenated and passed through additional fully-connected layers to predict drug response. Cancer cases were represented using gene expression profiles, while drugs were represented using molecular descriptors and fingerprints. To reduce the dimensionality of the gene expression input, only the LINCS L1000 landmark genes were included \cite{subramanian2017lincs}, which is a curated set of approximately 1,000 genes designed to capture a large proportion of transcriptomic variation. 

\subsubsection{1.2 IGTD}

Image Generator for Tabular Data (IGTD) \cite{zhu2021igtd} converts tabular features into two-dimensional image representations that can be processed by convolutional neural networks (CNNs). Features are assigned to pixel positions in images such that correlated features are positioned close to each other. In this study, IGTD was applied separately to cancer gene expression profiles and drug molecular descriptors, generating two image representations for each sample that were processed by separate CNN subnetworks. The resulting representations were concatenated and used to predict drug response. 

\subsubsection{1.3 GraphDRP}

GraphDRP \cite{Nguyen2021} is a graph-based deep learning model that represents drugs as molecular graphs, with atoms represented as nodes and chemical bonds represented as edges. Graph convolutional layers are used to learn drug representations directly from molecular structure. Cancer cases are represented using fixed-length molecular feature vectors. In this study, gene expressions were used as the cancer input. The learned drug and cancer representations are concatenated and passed through fully-connected layers to predict drug response. Although GraphDRP can incorporate additional omics feature types, including mutations and copy-numbers, only gene expressions were used in our analysis.

\subsubsection{1.4 HiDRA}

HiDRA (Hierarchical Network for Drug Response Prediction with Attention) \cite{Jin2021HiDRA} is an attention-based deep learning model that incorporates prior biological pathway information into the cancer representation. Gene expression features are first aggregated into pathway-level representations using gene-level attention. A second attention mechanism is then applied across pathways to generate an integrated representation. Drug structure is represented using 512-bit molecular fingerprints, and the drug and cancer representations are combined to predict drug response. The hierarchical attention architecture also provides a mechanism for identifying genes and pathways that contribute to model predictions. To improve computational efficiency, we modified the original HiDRA implementation by vectorizing the pathway-level computation. The original implementation processes each of the 186 pathways separately in a sequential order; in the modified implementation, pathway representations are combined into a single tensor so that pathway-level computations can be performed parallelly. This modification reduced model training time by more than 10 folds while producing highly consistent prediction outcomes. 

\subsubsection{1.5 LGBM}

Light Gradient-Boosting Machine (LGBM) \cite{lgbm} is a tree-based gradient boosting machine that constructs an ensemble of decision trees sequentially, with each tree trained to reduce the residual prediction error of the preceding ensemble. LGBM uses histogram-based feature binning and leaf-wise tree growth to efficiently model nonlinear relationships between input features and prediction target. Cancer cases were represented using gene expression profiles and drugs were represented using Mordred molecular descriptors. Models were trained for up to 10,000 boosting rounds with early stopping after 500 consecutive rounds without an improvement on validation performance. When training on CCL data, the maximum number of terminal leaves was set to 16, the maximum tree depth was set to 5, and the minimum number of samples per terminal leaf was set to 128 to limit model complexity and reduce overfitting.

\subsubsection{1.6 DeepTTC}

DeepTTC \cite{jiang2022deeptta} is a transformer-based deep learning model that predicts drug response using cancer gene expression profiles and drug substructure representations. Drug SMILES strings are decomposed into sequences of chemical substructures using an explainable substructure partition fingerprint method based on byte-pair encoding. These substructure sequences are then processed using a transformer encoder, allowing self-attention to model relationships among chemical substructures within each compound. Cancer gene expressions are encoded separately using a fully connected neural network. The resulting drug and cancer-sample representations are concatenated and passed through a final fully-connected prediction network to estimate drug response.

\subsection{2 Transfer Learning Implementation and Hyperparameter Optimization}
\label{sec:supp-transfer}

\subsubsection{2.1 Layer Freezing}


The model architecture of GraphDRP consists of three components: a drug encoder, a cancer encoder, and a prediction head that combines the cancer and drug embeddings to predict drug response. Transfer learning was implemented by freezing the early layers in the prediction head and updating its late layers using PDO data. To adjust the degree of adaptation during transfer learning, we varied the number of trainable layers within the prediction head. The drug and cancer encoders remained fixed during fine-tuning. The number of trainable layers was taken as a hyper-parameter and selected based on validation performance.

\subsubsection{2.2 Hyper-parameter Optimization}

Hyper-parameters involved in model fine-tuning were optimized independently for each transfer learning analysis using differential evolution (DE) \cite{Storn1997DifferentialEvolution, Virtanen2020SciPy}. DE is a population-based optimization algorithm in which individual members of the population represent candidate hyper-parameter configurations and fitness is determined by model performance on the validation set. Candidate configurations were iteratively generated through mutation and recombination, with configurations yielding improved validation performance retained for subsequent generations. Candidate solutions were initialized using Latin hypercube sampling to provide distributed coverage of the hyper-parameter search space. Mutation was performed using the \texttt{best1bin} strategy with mutation dithering over the interval $(0.5,\,1.0)$. The recombination probability was set to 0.7. For GraphDRP, the optimization jointly considered the number of trainable layers, learning rate, dropout rate, and batch size. Learning rate was optimized from $1 \times 10^{-6}$ to $1 \times 10^{-3}$ using logarithmic scaling. Dropout rate was optimized in the range of 0 to 0.90. Batch size was selected from $\{4, 8, 16\}$. DE sampled 20 candidate hyper-parameter configurations in each iteration and run for six optimization iterations, resulting in 120 candidate  configurations evaluated for each modeling analysis. 
For LGBM, optimization considered the learning rate and the minimum number of samples per leaf. Learning rate was optimized over the range 0.001--0.05, and the minimum number of samples per leaf was optimized between 2--64. DE sampled  10 candidate hyper-parameter configurations in each iteration and run for six optimization iterations, resulting in 60 candidate configurations evaluated per modeling analysis.

\subsection{3 Prediction Performance Measured by PCC}
\label{sec:supp-results}

\begin{table}[htbp]
\centering
\caption{Prediction performance (PCC, mean $\pm$ s.d.) of models trained exclusively on CCL data in random split cross-validations. CCL Test indicates the test set of CCL data. Prediction performance on CCL experiments and PDO experiments in IMPROVE datasets 1-3 are listed separately.}
\label{tab:phase1_pearson_random}
\resizebox{\linewidth}{!}{%
\begin{tabular}{lcccccccc}
\toprule
\multirow{2}{*}{\textbf{Model}} & \multirow{2}{*}{\textbf{CCL Test}} & \multicolumn{2}{c}{\textbf{IMPROVE Dataset 1}} & \multicolumn{2}{c}{\textbf{IMPROVE Dataset 2}} & \multicolumn{2}{c}{\textbf{IMPROVE Dataset 3}} & \multicolumn{1}{c}{\textbf{Pilot 1}} \\
\cmidrule(lr){3-4}\cmidrule(lr){5-6}\cmidrule(lr){7-8}\cmidrule(lr){9-9}
 & & CCL & PDO & CCL & PDO & CCL & PDO & PDO \\
\midrule
DeepTTC & 0.88$\,\pm\,$0.02 & 0.55$\,\pm\,$0.05 & 0.50$\,\pm\,$0.05 & 0.31$\,\pm\,$0.08 & 0.29$\,\pm\,$0.07 & 0.48$\,\pm\,$0.04 & 0.54$\,\pm\,$0.02 & 0.54$\,\pm\,$0.04 \\
UNO & 0.89$\,\pm\,$0.00 & 0.51$\,\pm\,$0.02 & 0.63$\,\pm\,$0.01 & 0.26$\,\pm\,$0.03 & 0.45$\,\pm\,$0.03 & 0.47$\,\pm\,$0.02 & 0.58$\,\pm\,$0.01 & 0.58$\,\pm\,$0.03 \\
GraphDRP & 0.89$\,\pm\,$0.00 & 0.55$\,\pm\,$0.02 & 0.58$\,\pm\,$0.02 & 0.28$\,\pm\,$0.04 & 0.43$\,\pm\,$0.03 & 0.48$\,\pm\,$0.03 & 0.55$\,\pm\,$0.02 & 0.56$\,\pm\,$0.02 \\
HiDRA & 0.88$\,\pm\,$0.00 & 0.59$\,\pm\,$0.05 & 0.49$\,\pm\,$0.05 & 0.30$\,\pm\,$0.08 & 0.31$\,\pm\,$0.05 & 0.55$\,\pm\,$0.05 & 0.49$\,\pm\,$0.06 & 0.56$\,\pm\,$0.03 \\
LGBM & 0.85$\,\pm\,$0.00 & 0.67$\,\pm\,$0.01 & 0.61$\,\pm\,$0.01 & 0.45$\,\pm\,$0.01 & 0.45$\,\pm\,$0.01 & 0.55$\,\pm\,$0.01 & 0.58$\,\pm\,$0.01 & 0.60$\,\pm\,$0.01 \\
IGTD & 0.88$\,\pm\,$0.00 & 0.54$\,\pm\,$0.03 & 0.60$\,\pm\,$0.03 & 0.28$\,\pm\,$0.04 & 0.46$\,\pm\,$0.04 & 0.46$\,\pm\,$0.03 & 0.53$\,\pm\,$0.03 & 0.58$\,\pm\,$0.03 \\
\midrule
\textit{Model Avg.} & 0.88$\,\pm\,$0.01 & 0.57$\,\pm\,$0.06 & 0.57$\,\pm\,$0.06 & 0.31$\,\pm\,$0.07 & 0.40$\,\pm\,$0.08 & 0.50$\,\pm\,$0.04 & 0.54$\,\pm\,$0.03 & 0.57$\,\pm\,$0.02 \\
\bottomrule
\end{tabular}%
}
\end{table}

\begin{table}[htbp]
\centering
\caption{Prediction performance (PCC, mean $\pm$ s.d.) of models trained exclusively on CCL data in cancer-blind cross-validations. CCL Test indicates the test set of CCL data. Prediction performance on CCL experiments and PDO experiments in IMPROVE datasets 1-3 are listed separately.}
\label{tab:phase1_pearson_cancerblind}
\resizebox{\linewidth}{!}{%
\begin{tabular}{lcccccccc}
\toprule
\multirow{2}{*}{\textbf{Model}} & \multirow{2}{*}{\textbf{CCL Test}} & \multicolumn{2}{c}{\textbf{IMPROVE Dataset 1}} & \multicolumn{2}{c}{\textbf{IMPROVE Dataset 2}} & \multicolumn{2}{c}{\textbf{IMPROVE Dataset 3}} & \multicolumn{1}{c}{\textbf{Pilot 1}} \\
\cmidrule(lr){3-4}\cmidrule(lr){5-6}\cmidrule(lr){7-8}\cmidrule(lr){9-9}
 & & CCL & PDO & CCL & PDO & CCL & PDO & PDO \\
\midrule
DeepTTC & 0.82$\,\pm\,$0.01 & 0.64$\,\pm\,$0.02 & 0.56$\,\pm\,$0.04 & 0.48$\,\pm\,$0.03 & 0.39$\,\pm\,$0.05 & 0.53$\,\pm\,$0.01 & 0.56$\,\pm\,$0.03 & 0.60$\,\pm\,$0.02 \\
UNO & 0.84$\,\pm\,$0.01 & 0.62$\,\pm\,$0.03 & 0.56$\,\pm\,$0.03 & 0.48$\,\pm\,$0.03 & 0.41$\,\pm\,$0.05 & 0.53$\,\pm\,$0.02 & 0.58$\,\pm\,$0.03 & 0.58$\,\pm\,$0.04 \\
GraphDRP & 0.83$\,\pm\,$0.01 & 0.60$\,\pm\,$0.02 & 0.57$\,\pm\,$0.03 & 0.41$\,\pm\,$0.05 & 0.44$\,\pm\,$0.03 & 0.55$\,\pm\,$0.02 & 0.54$\,\pm\,$0.02 & 0.54$\,\pm\,$0.02 \\
HiDRA & 0.83$\,\pm\,$0.01 & 0.65$\,\pm\,$0.02 & 0.53$\,\pm\,$0.04 & 0.43$\,\pm\,$0.09 & 0.34$\,\pm\,$0.07 & 0.54$\,\pm\,$0.02 & 0.54$\,\pm\,$0.04 & 0.60$\,\pm\,$0.02 \\
LGBM & 0.83$\,\pm\,$0.01 & 0.67$\,\pm\,$0.01 & 0.60$\,\pm\,$0.01 & 0.44$\,\pm\,$0.02 & 0.45$\,\pm\,$0.02 & 0.54$\,\pm\,$0.01 & 0.57$\,\pm\,$0.01 & 0.60$\,\pm\,$0.01 \\
IGTD & 0.83$\,\pm\,$0.01 & 0.65$\,\pm\,$0.02 & 0.61$\,\pm\,$0.03 & 0.45$\,\pm\,$0.04 & 0.46$\,\pm\,$0.03 & 0.53$\,\pm\,$0.03 & 0.55$\,\pm\,$0.03 & 0.60$\,\pm\,$0.02 \\
\midrule
\textit{Model Avg.} & 0.83$\,\pm\,$0.01 & 0.64$\,\pm\,$0.03 & 0.57$\,\pm\,$0.03 & 0.45$\,\pm\,$0.03 & 0.41$\,\pm\,$0.05 & 0.54$\,\pm\,$0.01 & 0.56$\,\pm\,$0.02 & 0.59$\,\pm\,$0.02 \\
\bottomrule
\end{tabular}%
}
\end{table}

\begin{table}[htbp]
\centering
\caption{Prediction performance (PCC, mean $\pm$ s.d.) of models trained using CCL and PDO data jointly in random split cross-validations. CCL Test indicates the test set of CCL data. Study CCL and Study PDO indicate the CCL experiments and the test set of PDO experiments in a PDO drug screening study, respectively.}
\label{tab:phase2_pearson_random}
\resizebox{\linewidth}{!}{%
\begin{tabular}{lccccccccccc}
\toprule
\multirow{2}{*}{\textbf{Model}} & \multicolumn{3}{c}{\textbf{IMPROVE Dataset 1}} & \multicolumn{3}{c}{\textbf{IMPROVE Dataset 2}} & \multicolumn{3}{c}{\textbf{IMPROVE Dataset 3}} & \multicolumn{2}{c}{\textbf{Pilot 1}} \\
\cmidrule(lr){2-4}\cmidrule(lr){5-7}\cmidrule(lr){8-10}\cmidrule(lr){11-12}
& CCL Test & Study CCL & Study PDO & CCL Test & Study CCL & Study PDO & CCL Test & Study CCL & Study PDO & CCL Test & Study PDO \\
\midrule
DeepTTC & 0.88$\,\pm\,$0.00 & 0.65$\,\pm\,$0.04 & 0.87$\,\pm\,$0.06 & 0.88$\,\pm\,$0.01 & 0.32$\,\pm\,$0.06 & 0.77$\,\pm\,$0.12 & 0.88$\,\pm\,$0.00 & 0.56$\,\pm\,$0.03 & 0.86$\,\pm\,$0.02 & 0.88$\,\pm\,$0.00 & 0.87$\,\pm\,$0.03 \\
UNO & 0.89$\,\pm\,$0.00 & 0.58$\,\pm\,$0.03 & 0.89$\,\pm\,$0.05 & 0.88$\,\pm\,$0.01 & 0.29$\,\pm\,$0.06 & 0.80$\,\pm\,$0.11 & 0.89$\,\pm\,$0.00 & 0.53$\,\pm\,$0.03 & 0.87$\,\pm\,$0.02 & 0.89$\,\pm\,$0.00 & 0.88$\,\pm\,$0.02 \\
IGTD & 0.88$\,\pm\,$0.01 & 0.61$\,\pm\,$0.04 & 0.84$\,\pm\,$0.07 & 0.87$\,\pm\,$0.01 & 0.32$\,\pm\,$0.06 & 0.76$\,\pm\,$0.14 & 0.88$\,\pm\,$0.00 & 0.56$\,\pm\,$0.04 & 0.85$\,\pm\,$0.03 & 0.88$\,\pm\,$0.00 & 0.86$\,\pm\,$0.03 \\
HiDRA & 0.88$\,\pm\,$0.00 & 0.65$\,\pm\,$0.02 & 0.89$\,\pm\,$0.03 & 0.88$\,\pm\,$0.00 & 0.29$\,\pm\,$0.03 & 0.78$\,\pm\,$0.13 & 0.88$\,\pm\,$0.00 & 0.60$\,\pm\,$0.01 & 0.85$\,\pm\,$0.02 & 0.88$\,\pm\,$0.00 & 0.85$\,\pm\,$0.02 \\
LGBM & 0.86$\,\pm\,$0.00 & 0.70$\,\pm\,$0.01 & 0.84$\,\pm\,$0.01 & 0.86$\,\pm\,$0.00 & 0.42$\,\pm\,$0.01 & 0.79$\,\pm\,$0.01 & 0.86$\,\pm\,$0.01 & 0.61$\,\pm\,$0.01 & 0.83$\,\pm\,$0.02 & 0.86$\,\pm\,$0.00 & 0.85$\,\pm\,$0.00 \\
GraphDRP & 0.88$\,\pm\,$0.01 & 0.64$\,\pm\,$0.03 & 0.83$\,\pm\,$0.07 & 0.87$\,\pm\,$0.01 & 0.34$\,\pm\,$0.05 & 0.76$\,\pm\,$0.12 & 0.88$\,\pm\,$0.00 & 0.55$\,\pm\,$0.03 & 0.83$\,\pm\,$0.03 & 0.88$\,\pm\,$0.00 & 0.86$\,\pm\,$0.03 \\
\midrule
\textit{Model Avg.} & 0.88$\,\pm\,$0.01 & 0.64$\,\pm\,$0.04 & 0.86$\,\pm\,$0.03 & 0.87$\,\pm\,$0.01 & 0.33$\,\pm\,$0.05 & 0.78$\,\pm\,$0.02 & 0.88$\,\pm\,$0.01 & 0.57$\,\pm\,$0.03 & 0.85$\,\pm\,$0.02 & 0.88$\,\pm\,$0.01 & 0.86$\,\pm\,$0.01 \\
\bottomrule
\end{tabular}%
}
\end{table}

\begin{table}[htbp]
\centering
\caption{Prediction performance (PCC, mean $\pm$ s.d.) of models trained using CCL and PDO data jointly in cancer-blind cross-validations. CCL Test indicates the test set of CCL data. Study CCL and Study PDO indicate the CCL experiments and the test set of PDO experiments in a PDO drug screening study, respectively.}
\label{tab:phase2_pearson_cancerblind}
\resizebox{\linewidth}{!}{%
\begin{tabular}{lccccccccccc}
\toprule
\multirow{2}{*}{\textbf{Model}} & \multicolumn{3}{c}{\textbf{IMPROVE Dataset 1}} & \multicolumn{3}{c}{\textbf{IMPROVE Dataset 2}} & \multicolumn{3}{c}{\textbf{IMPROVE Dataset 3}} & \multicolumn{2}{c}{\textbf{Pilot 1}} \\
\cmidrule(lr){2-4}\cmidrule(lr){5-7}\cmidrule(lr){8-10}\cmidrule(lr){11-12}
& CCL Test & Study CCL & Study PDO & CCL Test & Study CCL & Study PDO & CCL Test & Study CCL & Study PDO & CCL Test & Study PDO \\
\midrule
DeepTTC & 0.83$\,\pm\,$0.01 & 0.71$\,\pm\,$0.03 & 0.83$\,\pm\,$0.04 & 0.83$\,\pm\,$0.01 & 0.52$\,\pm\,$0.04 & 0.73$\,\pm\,$0.07 & 0.83$\,\pm\,$0.01 & 0.60$\,\pm\,$0.03 & 0.83$\,\pm\,$0.11 & 0.83$\,\pm\,$0.01 & 0.79$\,\pm\,$0.06 \\
UNO & 0.84$\,\pm\,$0.01 & 0.69$\,\pm\,$0.03 & 0.84$\,\pm\,$0.03 & 0.84$\,\pm\,$0.01 & 0.46$\,\pm\,$0.04 & 0.72$\,\pm\,$0.09 & 0.84$\,\pm\,$0.01 & 0.60$\,\pm\,$0.02 & 0.84$\,\pm\,$0.11 & 0.84$\,\pm\,$0.01 & 0.80$\,\pm\,$0.06 \\
IGTD & 0.82$\,\pm\,$0.01 & 0.70$\,\pm\,$0.04 & 0.74$\,\pm\,$0.05 & 0.82$\,\pm\,$0.01 & 0.44$\,\pm\,$0.08 & 0.69$\,\pm\,$0.11 & 0.83$\,\pm\,$0.01 & 0.60$\,\pm\,$0.04 & 0.81$\,\pm\,$0.11 & 0.83$\,\pm\,$0.01 & 0.78$\,\pm\,$0.05 \\
HiDRA & 0.83$\,\pm\,$0.00 & 0.72$\,\pm\,$0.01 & 0.81$\,\pm\,$0.03 & 0.83$\,\pm\,$0.00 & 0.37$\,\pm\,$0.01 & 0.72$\,\pm\,$0.04 & 0.83$\,\pm\,$0.00 & 0.61$\,\pm\,$0.01 & 0.82$\,\pm\,$0.12 & 0.83$\,\pm\,$0.00 & 0.77$\,\pm\,$0.04 \\
LGBM & 0.82$\,\pm\,$0.01 & 0.69$\,\pm\,$0.00 & 0.72$\,\pm\,$0.01 & 0.83$\,\pm\,$0.01 & 0.43$\,\pm\,$0.01 & 0.68$\,\pm\,$0.03 & 0.81$\,\pm\,$0.01 & 0.56$\,\pm\,$0.01 & 0.78$\,\pm\,$0.01 & 0.83$\,\pm\,$0.01 & 0.79$\,\pm\,$0.01 \\
GraphDRP & 0.83$\,\pm\,$0.01 & 0.69$\,\pm\,$0.03 & 0.79$\,\pm\,$0.04 & 0.82$\,\pm\,$0.02 & 0.42$\,\pm\,$0.04 & 0.72$\,\pm\,$0.09 & 0.83$\,\pm\,$0.01 & 0.62$\,\pm\,$0.03 & 0.80$\,\pm\,$0.12 & 0.83$\,\pm\,$0.01 & 0.78$\,\pm\,$0.06 \\
\midrule
\textit{Model Avg.} & 0.83$\,\pm\,$0.01 & 0.70$\,\pm\,$0.01 & 0.79$\,\pm\,$0.05 & 0.83$\,\pm\,$0.01 & 0.44$\,\pm\,$0.05 & 0.71$\,\pm\,$0.02 & 0.83$\,\pm\,$0.01 & 0.60$\,\pm\,$0.02 & 0.81$\,\pm\,$0.02 & 0.83$\,\pm\,$0.00 & 0.79$\,\pm\,$0.01 \\
\bottomrule
\end{tabular}%
}
\end{table}

\begin{table}[htbp]
\centering
\caption{Transfer learning prediction performance (PCC, mean $\pm$ s.d.) of models in random split cross-validations. CCL Test indicates the test set of CCL data. Study CCL and Study PDO indicate the CCL experiments and the test set of PDO experiments in a PDO drug screening study, respectively. Pilot 1 study does not include Study CCL experiments.}
\label{tab:phase3_pearson_random}
\resizebox{\linewidth}{!}{%
\begin{tabular}{lccccccccccc}
\toprule
\multirow{2}{*}{\textbf{Model}}
& \multicolumn{3}{c}{\textbf{IMPROVE Dataset 1}}
& \multicolumn{3}{c}{\textbf{IMPROVE Dataset 2}}
& \multicolumn{3}{c}{\textbf{IMPROVE Dataset 3}}
& \multicolumn{2}{c}{\textbf{Pilot 1}} \\
\cmidrule(lr){2-4}
\cmidrule(lr){5-7}
\cmidrule(lr){8-10}
\cmidrule(lr){11-12}
& CCL Test & Study CCL & Study PDO
& CCL Test & Study CCL & Study PDO
& CCL Test & Study CCL & Study PDO
& CCL Test & Study PDO \\
\midrule
GraphDRP
& 0.70$\,\pm\,$0.18 & 0.76$\,\pm\,$0.04 & 0.90$\,\pm\,$0.03
& 0.75$\,\pm\,$0.05 & 0.48$\,\pm\,$0.13 & 0.78$\,\pm\,$0.14
& 0.77$\,\pm\,$0.05 & 0.61$\,\pm\,$0.04 & 0.86$\,\pm\,$0.03
& 0.71$\,\pm\,$0.05 & 0.87$\,\pm\,$0.02 \\

LGBM
& 0.62$\,\pm\,$0.04 & 0.81$\,\pm\,$0.01 & 0.87$\,\pm\,$0.05
& 0.71$\,\pm\,$0.01 & 0.62$\,\pm\,$0.04 & 0.82$\,\pm\,$0.10
& 0.73$\,\pm\,$0.03 & 0.58$\,\pm\,$0.02 & 0.86$\,\pm\,$0.03
& 0.71$\,\pm\,$0.03 & 0.87$\,\pm\,$0.02 \\
\bottomrule
\end{tabular}%
}
\end{table}


\begin{table}[htbp]
\centering
\caption{Transfer learning prediction performance (PCC, mean $\pm$ s.d.) of models in cancer-blind cross-validations. CCL Test indicates the test set of CCL data. Study CCL and Study PDO indicate the CCL experiments and the test set of PDO experiments in a PDO drug screening study, respectively. Pilot 1 study does not include Study CCL experiments.}
\label{tab:phase3_pearson_cancerblind}
\resizebox{\linewidth}{!}{%
\begin{tabular}{lccccccccccc}
\toprule
\multirow{2}{*}{\textbf{Model}}
& \multicolumn{3}{c}{\textbf{IMPROVE Dataset 1}}
& \multicolumn{3}{c}{\textbf{IMPROVE Dataset 2}}
& \multicolumn{3}{c}{\textbf{IMPROVE Dataset 3}}
& \multicolumn{2}{c}{\textbf{Pilot 1}} \\
\cmidrule(lr){2-4}
\cmidrule(lr){5-7}
\cmidrule(lr){8-10}
\cmidrule(lr){11-12}
& CCL Test & Study CCL & Study PDO
& CCL Test & Study CCL & Study PDO
& CCL Test & Study CCL & Study PDO
& CCL Test & Study PDO \\
\midrule
GraphDRP
& 0.70$\,\pm\,$0.09 & 0.78$\,\pm\,$0.05 & 0.85$\,\pm\,$0.05
& 0.68$\,\pm\,$0.07 & 0.64$\,\pm\,$0.08 & 0.79$\,\pm\,$0.06
& 0.66$\,\pm\,$0.08 & 0.59$\,\pm\,$0.03 & 0.83$\,\pm\,$0.11
& 0.68$\,\pm\,$0.10 & 0.78$\,\pm\,$0.06 \\

LGBM
& 0.68$\,\pm\,$0.08 & 0.80$\,\pm\,$0.03 & 0.84$\,\pm\,$0.03
& 0.73$\,\pm\,$0.04 & 0.57$\,\pm\,$0.06 & 0.78$\,\pm\,$0.06
& 0.70$\,\pm\,$0.04 & 0.59$\,\pm\,$0.02 & 0.82$\,\pm\,$0.13
& 0.68$\,\pm\,$0.03 & 0.81$\,\pm\,$0.05 \\
\bottomrule
\end{tabular}%
}
\end{table}

\end{document}